\documentclass[12pt]{article}
\usepackage[mathscr]{eucal}
\usepackage{epsfig,amsfonts}
\usepackage[fleqn]{amsmath}
\usepackage{amsthm,amssymb}
\usepackage{graphicx}
\usepackage{hhline}
\usepackage{cite}

\usepackage[mathscr]{eucal}
\usepackage{epsfig,amsfonts}
\usepackage[fleqn]{amsmath}
\usepackage{amsthm,amssymb}
\usepackage{graphicx}
\usepackage{hhline}
\usepackage{cite}
\usepackage{psfrag}
\usepackage[dvips]{color}
\usepackage{graphicx,amsmath,amsfonts}
\usepackage{epsfig,amsfonts}
\usepackage[mathscr]{eucal}
\usepackage{epsfig,amsfonts}

\makeatletter
\@addtoreset{equation}{section}
\makeatother

\def\be{\begin{equation}}
\def\ee{\end{equation}}
\def\bdm{\begin{displaymath}}
\def\edm{\end{displaymath}}
\def\bea{\begin{eqnarray}}
\def\eea{\end{eqnarray}}

\def\zb{{\bar z}}

\def\ri{{\rm i}}

\def\XXint#1#2#3{{\setbox0=\hbox{$#1{#2#3}{\int}$}
    \vcenter{\hbox{$#2#3$}}\kern-.5\wd0}}

\newcommand{\rd}{\mbox{d}}
\newcommand{\re}{\mbox{e}}

\begin{document}

\begin{titlepage}
\begin{flushright}
RUNHETC-2013-06\\
\end{flushright}

\begin{center}
\begin{LARGE}

{\bf ODE/IM correspondence for the Fateev model}

\end{LARGE}
\vspace{1.3cm}

\begin{large}

{\bf  Sergei  L. Lukyanov}$^{1,2}$

\end{large}

\vspace{1.cm}
${}^{1}$NHETC, Department of Physics and Astronomy\\
     Rutgers University\\
     Piscataway, NJ 08855-0849, USA\\
\vspace{.2cm}
and\\
\vspace{.2cm}
${}^{2}$L.D. Landau Institute for Theoretical Physics\\
  Chernogolovka, 142432, Russia\\
\vspace{1.0cm}

\end{center}

\begin{center}

\vspace*{0.1in} \begin{center} {\em  Dedicated to my PhD advisor V.A. Fateev on the occasion of his 65th anniversary}
\end{center}

\vspace{.6cm}

\centerline{\bf Abstract} \vspace{.8cm}

\parbox{13cm}{The Fateev model is somewhat special among two-dimensional quantum field theories.
For different values of the parameters,
it can be reduced to  a variety of integrable systems.
An incomplete list of the reductions
includes   $O(3)$ and $O(4)$ non-linear sigma models and their continuous deformations (2D and 3D sausages,
anisotropic principal chiral field),  the Bukhvostov-Lipatov model, the
$N=2$ supersymmetric sine-Gordon model, as well as the
integrable perturbed $SU_2(n)\otimes SU_2(p-2)/SU_2(n+p-2)$ coset CFT.
The model   possesses  a  mysterious  symmetry structure of
the exceptional quantum superalgebras $U_q{\widehat{\big(D(2|1;\alpha)}}\big)$.\\
In this work, we propose the ODE/IM correspondence between  the Fateev model and a
certain generalization of the classical  problem of constant mean curvature embedding
of a thrice-punctured  sphere   in $AdS_3$.}

\end{center}

\vspace{.2cm}
\begin{flushleft}
\rule{3.1 in}{.007 in}\\
{March 2013}

\end{flushleft}

\vfill

\end{titlepage}

\newpage

\section{Introduction}

Broadly  speaking the ODE/IM correspondence is a link 
between 
the theory of  {\it Integrable Models} in two dimensions and the  spectral analysis of
{\it Ordinary Differential Equations}.
The approach was originated by  Dorey and Tateo  in Ref.\cite{Dorey:1998pt}
from the  observation  that
the thermodynamic Bethe ansatz equations for certain 2D CFT minimal models    proposed   in \cite{Bazhanov:1994ft}
coincide
with an  exact version of the Bohr-Sommerfeld quantization  condition 
for 1D anharmonic  oscillator -- a  remarkable result  
due to  
Voros \cite{Voros:1992, Voros:1994, Voros:1999bz}.
Very shortly the initial  observation    was  generalized and   proved in  Ref.\cite{Bazhanov:1998wj}.
Later the   ODE/IM correspondence was established    for a large  variety of    models  (for review see \cite{Dorey:2007zx}).
However, during the next decade, all attempts  to incorporate
massive  integrable QFT  in the ODE/IM   correspondence have  failed.
Since   the work of Gaiotto, Moore and Neitzke \cite{Gaiotto:2008cd},
thermodynamic Bethe ansatz  equations have  emerged in   different contexts  of  SUSY gauge theories,
algebraic and differential geometry \cite{Gaiotto:2009hg, Alday:2009yn, Alday:2009dv, Alday:2010vh}.
Inspired by this  progress
Zamolodchikov and the author 
have
established the ODE/IM correspondence for
the quantum  sine/sinh-Gordon model  -- the model which  
always served  as a basis for the development 
of  the 2D integrable   QFT  \cite{Lukyanov:2010rn}.
Recently the result of Ref.\cite{Lukyanov:2010rn} was extended to  the Toda type QFT model\ \cite{Dorey:2012bx}.
It has  become   clear  that the ODE/IM correspondence is
one of the most important    aspects
of integrability in 2D QFT. 
Among  open problems within the approach
is how to incorporate
the class of integrable non linear sigma models  into the  ODE/IM correspondence.
In this work, we propose an example  which aims  to step in this direction.

We begin with the following well known fact (see e.g. Ref.\cite{Troyanov});
\vskip 0.2cm
\noindent
Let $\Sigma_{g,n}$ be a compact  Riemann surface with $n$ marked points (``punctures'')
and $a_1,\, a_2,\ldots a_n$  be positive
numbers such that
$2\, \chi(\Sigma_{g})+\sum_{i=1}^n(a_i-2)=0$.
Then there exists a flat metric on $\Sigma_{g,n}$ with conical singularities of angle $\pi a_i$ at
the ${ i}^{{\rm th}}$ puncture.
The metric is unique up to homothety.

\vskip 0.2cm
In the case 
 of a
two-sphere  with three  punctures
the theorem's   condition reads as
\bea\label{aopsaosap}
a_1+a_2+a_3=2\ ,
\eea
whereas  its  statement is   somewhat trivial.
Indeed introduce  a complex coordinate $z$ and  define
a holomorphic differential $p(z)\,(\rd z)^2$
on the  universal cover of $\Sigma_{0,3}$
by means of  the following 
assignment for     $p(z)$:
\bea\label{sossao}
p(z)=\rho^2\ \frac{(z_3-z_2)^{a_1}\,(z_1-z_3)^{a_2}\,(z_2-z_1)^{a_3}}
{(z-z_1)^{2-a_1}(z-z_2)^{2-a_2}(z-z_3)^{2-a_3}}\ .
\eea
Then the  flat  metric reads explicitly
\bea\label{oiaoisao}
(\rd s)^2_0=\sqrt{p(z){\bar p}({\bar z})}\ {\rd z}{\rd {\bar z}}\ .
\eea
Here 
$\rho$ stands for   the  homothety   parameter and
$z_i$ labels the punctures.

Consider now the problem of   constant  
mean curvature    embedding of  $\Sigma_{0,3}$ into   $AdS_3$.
In this case, the Gauss-Peterson-Codazzi equation can be brought to the form of 
the modified  Sinh-Gordon (MShG)  equation 
\bea\label{akasjksa}
\partial_z\partial_{\bar z}{ \eta}-\re^{2{ \eta}}+ p(z){\bar p}({\bar z})\, \re^{-2{ \eta}}=0\ ,
\eea
where  the field $\eta$ defines   the induced metric \cite{Pohlmeyer:1975nb,Bobenko, DeVega:1992xc,Alday:2009yn}\footnote{The derivation 
of \eqref{akasjksa}  is nearly   identical to  the ones from  textbook examples of
embedding  into ${\mathbb R}^3$, ${\mathbb S}^3$ or  ${\mathbb H}^3$
(see, e.g. Ref.\cite{Bobenko}).
However in that cases Eqs.\eqref{akasjksa} 
is substituted by  $\partial_z\partial_{\bar z}{ \eta}+\re^{2{ \eta}}- p(z){\bar p}({\bar z})\, \re^{-2{ \eta}}=0$. }
\bea\label{sissu}
(\rd s)^2_{\rm cmc}=\frac{4}{1+H^2}\  \frac{\re^{2\eta}}{ \sqrt{p(z){\bar p}({\bar z})}}\  (\rd s)^2_0\ ,
\eea
and $H=const$ stands for the mean curvature.
A suitable   solution  should be    real and smooth as $z\not= z_i$, and, if  we want to preserve the  amount of 
the Gaussian curvature localized at the punctures,  it should satisfy the
conditions
\bea\label{asosospad}
\eta- {\textstyle \frac{1}{4}}\ \log\big(\,p(z){\bar p}({\bar z})\,)=O(1)\ \ \ \ \ \ {\rm at}\ \ \ z\to z_i\ \ (i=1,2,3) \ \ {\rm and} \ \infty\  .
\eea

It turns out that the problem  \eqref{akasjksa},\,\eqref{asosospad}
admits an important generalization. 
One can   consider the MShG equation in  the  flat background of the  punctured sphere  $\Sigma_{0,3}$
subject of 
more general  asymptotic
conditions 
\bea\label{asosospa}
\eta&=&-2\, \log|z|+O(1)\ \ \ \ \ \ \ \ \ \ \ \  \ \ \ {\rm at}\ \ \ \ \ \  z\to\infty \nonumber\\
\eta&=& 2m_i\, \log|z-z_i|+O(1)\ \ \ \ \ \ \ \ {\rm at}\ \ \ \ \ \   z\to z_i\ .
\eea
If
\bea\label{posaposapo}
0<a_i<2
\eea
and 
\bea\label{ospssaps}
-\frac{1}{2}< m_i\leq -\frac{1}{4}\ (2-a_i)\ ,
\eea
then the  solution of   the  generalized   problem  exists and  is unique.
We are not going to prove this statement here.
Instead,
we will accept it and  argue  that  the solution of 
\eqref{akasjksa},\,\eqref{asosospa}
is   related to a certain 2D integrable QFT model, where
all the parameters  $a_i$,\ $\rho$ and  $m_i$, as well as the restrictions
\eqref{posaposapo} and \eqref{ospssaps}, admit  simple physical interpretations.

\begin{figure}
\centering
\includegraphics[width=6.  cm]{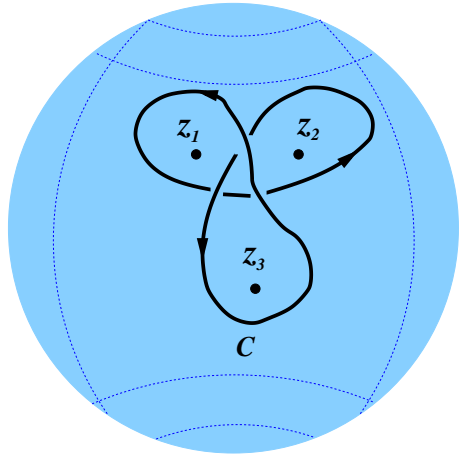}
\caption{The integration contour $C$  for the conserved charges  ${\mathfrak q}_{2n-1}$.}
\label{fig1aa}
\end{figure}

To describe the relation between classical and quantum systems
we shall need an important property of the problem\ \eqref{akasjksa},\,\eqref{asosospa}.
The essential point in the
formal theory of the  equation \eqref{akasjksa}  is an existence of
an infinite  hierarchy of  one-forms, which are closed by virtue of the MShG equation only,
\bea\label{spsapsaos}
\{\omega_{2n}\}_{n=1}^\infty\ :\ \ \ \ \ \ \ \ \ \ \ \ \ \rd \omega_{2n}=0\  .
\eea
They  are usually normalized by 
the condition
\bea\label{aopssap}
\omega_{2n}=
\Big(\, \big(\sqrt{p(z)}\,\big)^{1-2n}\,(\partial_z\eta)^{2n}+\ldots\,\Big)\, \rd z+
\big(\,\ldots\,\big)\ \rd {\bar z}\ ,
\eea
where dots in the first bracket involves terms with  higher derivatives of $\partial_z\eta$  and/or $p(z)$. 
The one-forms are not single-valued on the punctured sphere 
and should be considered on the  universal cover of $\Sigma_{0,3}$ 
caused by the presence of  the multivalued function  $\sqrt{p(z)}$.
However,    the contour $C$
depicted in   Fig.\ref{fig1aa}, loops around each puncture  and  $\sqrt{p(z)}$ acquires  the same value
after the analytic   continuation along $C$.
Therefore
the restriction of $\omega_{2n}$ to $C$ are  single-valued and  the integrals
\bea\label{osapsospa}
{\mathfrak q}_{2n-1}=\oint_{C}\omega_{2n}
\eea
are  not sensitive to continuous deformations of the contour.
A branch of multivalued function $\sqrt{p(z)}$  can
be chosen in such a way that the conserved charges     are  real numbers,
\bea\label{spsaopsaosa}
{\mathfrak q}_{2n-1}={\mathfrak q}^*_{2n-1}\ .
\eea

We now turn to the  QFT model of our interest.
It was  introduced by  Fateev in Ref.\cite{Fateev:1996ea}
and governed by
the  following  Lagrangian in the $1+1$ Minkowski
space
\bea\label{aposoasio}
{\cal L}&=& \frac{1}{16\pi}\ \sum_{i=1}^3
\big(\, (\partial_t\varphi_i)^2-(\partial_x\varphi_i)^2\,\big)\\
&+&2\mu\
\big(\, \re^{\ri\, \alpha_3\varphi_3}\ \cos(\alpha_1\varphi_1+\alpha_2\varphi_2)+\re^{-\ri \alpha_3\varphi_3}\
\cos(\alpha_1\varphi_1-\alpha_2\varphi_2)\,\big)\,  .
\nonumber
\eea
Here $\alpha_i$ are    coupling constants subject to a single constraint
\bea\label{aposapoas}
\alpha_1^2+\alpha_2^2+\alpha_3^2=\frac{1}{2}\ .
\eea
In this paper, we shall focus on the case where
\bea\label{aopssaspsopa}
\alpha_1^2>0\,,\ \ \ \ \ \ \ \alpha_2^2>0\,,\ \ \ \ \ \ \
\alpha_3^2>0\,\ .
\eea
The parameter $\mu$ in the Lagrangian sets the mass scale, $\mu\sim [\,{\rm mass}\,]$.
We shall consider  the theory in finite-size geometry, with the
spatial coordinate $x$ in $\varphi_i=\varphi_i(x,t)$ compactified on a circle of
circumference $R$, with the periodic boundary conditions
\bea\label{sissiaosai}
\varphi_i(x+R,t)=\varphi_i(x,t)\ .
\eea
Due to the periodicity of the potential term in  \eqref{aposoasio} in $\varphi_i$,  the space of states ${\cal H}$ splits on the orthogonal
subspaces ${\cal H}_{k_1,k_2,k_3}$ characterized by the three  ``quasimomentums'' $k_i$:
\bea\label{sapsapo}
\varphi_i\to\varphi_i+2\pi/\alpha_i\ :\ \ \ |\,\Psi_{k_1,k_2,k_3}\,\rangle\to\re^{2\pi\ri k_i}\  |\,\Psi_{k_1,k_2,k_3}\,\rangle\ .
\eea

The QFT \eqref{aposoasio} is integrable, in particular
it has infinite set of commuting local integrals of motion 
$\mathbb{I}^{(+)}_{2n-1}$,\ $\mathbb{ I}^{(-)}_{2n-1}$, $2n=2,\, 4,\,6,\,\ldots$
being the Lorentz spins of the associated local densities \cite{Fateev:1996ea}:
\bea\label{isusospsasopas}
\mathbb{I}^{(\pm)}_{2n-1}=
\int_0^R\frac{\rd x}{ 2\pi}\
\ \Big[\, \sum_{i+j+k=n} C^{(n)}_{ijk}\ (\partial_\pm\varphi_1)^{2i}\,
(\partial_\pm\varphi_2)^{2j}\, (\partial_\pm\varphi_3)^{2k} +\ldots\,  \Big]\ ,
\eea
where  $\partial_\pm=\frac{1}{2} (\partial_x\mp \partial_t)$ 
and
$\ldots$ stand for the terms involving higher
derivatives of $\varphi_i$, as well as the terms proportional to
powers of $\mu$.
The constant $C^{(n)}_{ijk}$ was found in Ref.\cite{Lukyanov:2012wq}
\bea\label{saosopsaosa}
C^{(n)}_{ijk}=\frac{n!}{i!\ j!\ k!}\  \ 
\frac{
\big(2\alpha_1^2(1-2 n)\big)_{n-i} \big(2\alpha_2^2\,(1-2 n)\big)_{n-j}\,\big(2\alpha_3^2\,(1-2 n)\big)_{n-k}}{
(2n-1)^3\  (4\alpha^2_1)^{1-i}\ (4\alpha^2_2)^{1-j}\ (4\alpha^2_3)^{1-k}}\ , 
\eea
where   $(x)_n$ stands for the Pochhammer symbol.
Notice that
the displayed terms in \eqref{isusospsasopas} with the given $C^{(n)}_{ijk}$
set   the normalization of $\mathbb{I}^{(\pm)}_{2n-1}$ unambiguously.

Of primary interest are the $k$-vacuum
eigenvalues
\bea\label{akssasuaias}
I_{2n-1}=I^{(+)}_{2n-1}(\{k_i\}\,|\,R)={ I}^{(-)}_{2n-1}(\{k_i\}\,|\,R)\ ,
\eea
especially
the $k$-vacuum energy
\bea\label{apospsapo}
E=2\ I_1\ .
\eea
In the large-$R$ limit all vacuum  eigenvalues $I_{2n-1}$
vanish except $I_1$. 
The vacuum energy is composed of an extensive part proportional to the length of
the system,
\bea\label{saospsaos}
E=R\,{\cal E}_0+o(1)\ \ \ \ \ \ \ \ \ {\rm at }\ \ \ R\to \infty\ .
\eea
One of  Fateev's impressive  results concerning  
theory \eqref{aposoasio} is an elegant  analytical expression for the
specific  bulk energy
\bea\label{asososa}
{\cal E}_0=-\pi\mu^2\ \prod_{i=1}^3\frac{\Gamma(2\alpha_i^2)}{\Gamma(1-2\alpha_i^2)}\ .
\eea

The main  observation of this work is  that  the  vacuum eigenvalues 
can be expressed in terms of  the classical  conserved charges \eqref{osapsospa}:
\bea\label{aspspsapo}
\mu^{-1}\ \big(\,I_{1}-{\textstyle\frac{1}{2}}\, R\,{\cal E}_0\,\big)&=&d_1\ {\mathfrak q}_{1}\\
\mu^{1-2n}\ I_{2n-1}&=&d_n\ {\mathfrak q}_{2 n-1}\ \  \ \ \ \ \ \ \ \ \ \ \ \ (n=2,\,3,\,\ldots)\  .\nonumber
\eea
Here
$d_n$ are  constants, independent of $k_i$ and $R$.
  With the normalization conditions for ${\mathfrak q}_{2n-1}$ and
${\mathbb I}^{(\pm)}_{2n-1}$ described above, $d_n$   reads explicitly as
\bea\label{poapsospsaos}
d_n=(2\pi)^{2n-1}\ 
\ \frac{(-1)^{n-1} }{16\,\pi^2  }\  
\ \prod_{i=1}^3\Gamma\big(\, 2\,(2n-1)\, \alpha^2_i\,\big) \ .
\eea
The parameters of the quantum and classical problems are identified as follows:
\bea\label{sssaopsa}
\alpha_i^2&=&{\frac{a_i}{4}}\ \ \ \ \ \ \ \  \ (i=1,2,3)\nonumber\\
|k_i|&=&\frac{1}{a_i}\ (2m_i+1)\ ,
\eea
whereas the relation between dimensionless parameter $\mu R$ and $\rho$ is  given by 
\bea\label{saossaops}
\mu R=2 \rho\ .
\eea

Notice that the quantum  integrals of motion $\mathbb{I}^{(\pm)}_{2n-1}$ are  periodic functions of the quasimomentums, and
one can assume
that  $k_i$ takes values within the
first Brillouin zone
\bea\label{sosasoias}
-\frac{1}{2}<k_i\leq\frac{1}{2}\ .
\eea
Since  $\mathbb{I}^{(\pm)}_{2n-1}$
commute with the ``charge conjugations'' --  the obvious  symmetries  $\varphi_i\to-\varphi_i\ (i=1, 2, 3)$
of the Lagrangian, the
eigenvalues \eqref{akssasuaias} are even functions of $k_i$.
In a view of the  identification \eqref{sssaopsa},
the inequality\ \eqref{sosasoias} 
suggests a natural domain \eqref{ospssaps} for the parameters  $m_i$.
Although 
both the classical conserved charges  and the eigenvalues $I_{2n-1}$  are nonsingular at  $k_i=0\  (m_i=-\frac{1}{2}$), 
the  small-$|z- z_i|$  asymptotic \eqref{asosospa}
contains a 
subleading $\log\log$-term in this case (the so-called ``parabolic point'').  For this reason,
the value $m_i=-\frac{1}{2}$ is excluded from the domain   \eqref{ospssaps}.

Let us recall  now
that  the  MShG equation
constitutes a flatness condition
for  a certain   $sl(2)$-valued connection ${\boldsymbol A}={\boldsymbol A}_z\rd z+ {\boldsymbol A}_{\bar z}\rd {\bar z}$.
The associated linear problem
\bea\label{lp}
(\partial_z-{\boldsymbol A}_z)\, {\boldsymbol\Psi}=0\ ,\ \ \ \ \ \ \ \ \ \ \ (\partial_{\bar z}- {\boldsymbol A}_{\bar z})\, 
{\boldsymbol\Psi}=0
\eea
is of prime importance for solving
\eqref{akasjksa},\,\eqref{asosospa}.
In particular, certain  monodromy  coefficients for  the linear system \eqref{lp}
can be regarded  as  generating functions for the set  of conserved charges
$\{{\mathfrak q}_{2n-1}\}_{n=1}^\infty$.
Thus  Eqs.\eqref{aspspsapo} relate     spectral characteristics  of  the  linear problem \eqref{lp}
with  a  vacuum  spectrum of  the local  integral of motions from the Fateev model.
This is an example of  the   ODE/IM correspondence.\footnote{
In fact,  the abbreviation ``IM''  can be understood  as  a shortened form of  either {\it Integrable Models} or 
{\it Integrals of Motion}.
The author   slants toward  the second interpretation; Until now  there is no any indication that the correspondence
can be extended beyond the scope of relations between  the spectral characteristics. In particular,
the relation of  the formal variables $(z,{\bar z})$ and the space-time coordinates $(x,t)$,
as well as the
r${\hat {\rm o}}$le of the classical field $\eta$ itself in the quantum theory,
remains completely  mysterious.}

Al.B. Zamolodchikov was probably the first who realized the main advantages of ODE/IM correspondence
compare to the traditional approaches \cite{AlZ};
The ODE counterpart makes 
explicit   the   analytic properties    of  relevant physical quantities
considered as  functions of  the  parameters. Within the  ODE/IM approach,  the integrable  model can be studied uniformly
in different parameter regimes.
In the case of  massive QFT,
this   was explicitly  demonstrated  in Ref.\cite{Lukyanov:2010rn}
for   the quantum sine- and sinh-Gordon models.
In spite of formal similarity of the Lagrangians, the physical content of these models is  
very different from one another. 
However, the  models can be treated uniformly 
within the  ODE/IM approach.
With only a few minor modifications
one can jump  from the trigonometric to the hyperbolic model.
The story is repeating itself --
only  little modifications $\grave{\rm a}$ la
sin/sinh-Gordon one are  required to 
extend 
the above  ODE/IM  correspondence   to the most interesting  regime of QFT \eqref{aposoasio}
with
\bea\label{ussyysaopssaspsopa}
\alpha_1^2>0\,,\ \ \ \ \ \ \ \alpha_2^2>0\,,\ \ \ \ \ \ \
\alpha_3^2<0\, .
\eea
In this  regime, the  Fateev model
provides a dual description of the  ``3D-sausage'' --  the integrable non-linear sigma model with
three dimensional target space \cite{Fateev:1996ea}. The sigma-model regime will be the subject of a separate publication.
The main purpose of this work is to present  evidences in support of the relations \eqref{aspspsapo}-\eqref{saossaops} in the
regime \eqref{aopssaspsopa}.

The paper is organized as follows. In Section\,\ref{defin} we give an accurate definition of
the conserved charges ${\mathfrak q}_{2n-1}$ and
recall their relation to  the
flat 
connection.
In the next section we focus on the first conserved charge ${\mathfrak q}_{1}$.
Our
analysis is  based on the observation that as $\rho\to 0$
the problem  \eqref{akasjksa},\,\eqref{asosospa}
is reduced to the problem of  constructing  a  metric of  constant intrinsic curvature
on the punctured sphere.  Using the results from the classical Liouville 
theory we derive the small-$\rho$ expansion of ${\mathfrak q}_{1}$. 
Section\,\ref{qn} deals with  the $\rho\to0$ limit of the conserved charges ${\mathfrak q}_{2n-1}$ with $n>1$.
Finally, in Section\,\ref{id} we extract the relations
\eqref{aspspsapo}-\eqref{saossaops}
from the comparison of
the    small-$\rho$  expansions for  ${\mathfrak q}_{2n-1}$ 
against   
predictions of the Conformal Perturbation Theory for
the QFT  \eqref{aposoasio}.
We conclude with a few remarks on a rigorous proof
of the proposed ODE/IM correspondence.

In the theory of integrable models,  an important r${\hat {\rm o}}$le 
belongs to the  concept of the Yang-Yang function(al) \cite{Nekrasov:2009rc}.
It is  of prime interest to  the ODE/IM  correspondence \cite{Lukyanov:2011wd}.
In this work, we  use  the  Yang-Yang function  as an auxiliary  tool only
to derive the small $\rho$-expansion of the conserved charge ${\mathfrak q}_1$.
Because of its own significance,
we present  some  facts  concerning  the  Yang-Yang function for the Fateev model  in  the appendix.

\section{\label{defin}Conserved charges for MShG on the punctured sphere}

In this section  we give detailed description of the conserved charges $\{{\mathfrak q}_{2n-1}\}_{n=1}^\infty$. 

\subsection{From MShG to ShG}

\begin{figure}
\centering
\includegraphics[width=7  cm]{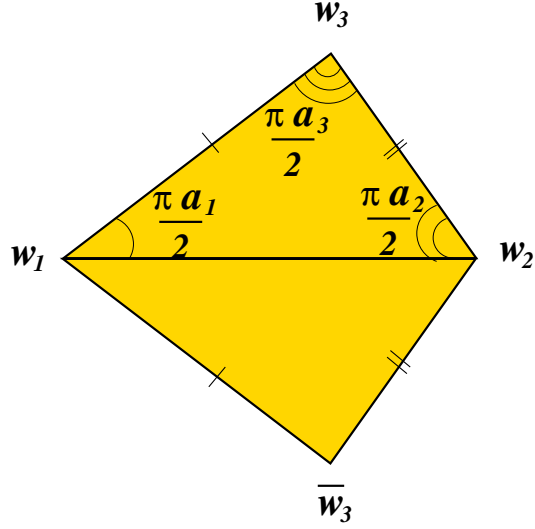}
\caption{Triangle $(w_1, w_2, w_3)$ is a $w$-image of the 
upper half plane $\Im m(\zeta)>0$ under the  Schwarz-Christoffel mapping \eqref{soiosa}. 
The point ${\bar w}_3$ is a reflection of $ w_3$ w.r.t. the straight line $(w_1,w_2)$. The domain  ${\mathbb D}$ is obtained
from the 4-polygon  $(w_1, w_3, w_2,{\bar w}_3)$
by  the  identification
of the sides $[w_1,w_3]$,  $[w_2,w_3]$ and $[w_1,{\bar w}_3]$,  $[w_2,{\bar w}_3]$, respectively.}
\label{fig8}
\end{figure}

The flat metric defines a complex structure
on a punctured sphere  $\Sigma_{0,3}$. Since the MShG equation \eqref{akasjksa}
is covariant under coordinate diffeomorphisms,
its form  can be simplified  by
taking  advantage of
conformal transformations.
First of all,  we can  use  the  M\"obius  transformation
\bea\label{sospsaos}
\zeta=
\frac{(z_2-z_3)(z-z_1)}{(z_2-z_1)(z-z_3)}
\eea
to 
convert the puncture's coordinates   $(z_1, z_2, z_3)$ to their standard values  $(0, 1, \infty)$. This
brings  Eq.\eqref{akasjksa} to the form
\bea\label{isisiosa}
\partial_\zeta\partial_{\bar \zeta}{\tilde \eta}-\re^{2{ \tilde \eta}}+ {P}(\zeta){\bar {P}}({\bar \zeta})\, \re^{-2{ \tilde \eta}}=0\ ,
\eea
where 
\bea\label{saosposa}
{\tilde \eta}={ \eta}+\frac{1}{2}\ \log \Big(\frac{\rd z}{\rd \zeta}\, \frac{\rd {\bar z}}{\rd {\bar \zeta}}\Big)\ ,\ \ \ \ \ \ 
{P}(\zeta)=\rho^2\   \zeta^{a_1-2}\ (1-\zeta)^{a_2-2} \ .
\eea
Then we
consider  the  Schwarz-Christoffel mapping
\bea\label{soiosa}
w(\zeta)=\int\rd \zeta\ \sqrt{P(\zeta)}\ ,
\eea
which transforms the upper half plane $\Im  m(\zeta)\geq 0$ to the triangle $(w_1,\,w_2,\,w_3)$ in the complex $w$-plane
(see Fig.\,\ref{fig8}).
The lower half plane $\Im  m(\zeta)\leq 0$ can be   mapped into the triangle reflected w.r.t. the  straight line $(w_1,w_2)$.
If there exists a   unique  solution, it    should respect all
symmetries of
the problem under consideration. In particular it should be invariant under
the reflection
$\zeta\leftrightarrow {\bar \zeta}$:
\bea\label{soosaos}
{\tilde \eta}(\zeta,{\bar \zeta})=
{\tilde \eta}({\bar \zeta},\zeta)\ .
\eea
Let  ${\mathbb D}$ be a 
four-sided polygon  with the vertexes located at $w_i=\{w_1,\, w_2,\, w_3,\, {\bar w}_3\}$  whose adjacent sides are  glued together
as it is  shown in  Fig.\,\ref{fig8}
to form a  topological two-sphere.
The overall scale of the polygon is governed by the homothety parameter $\rho$  in Eq.\eqref{sossao}.
Without loss of generality we shall assume that  $\rho>0$, then
\bea\label{soosapsosa}
|w_i-w_j|
=\rho \ \ \frac{\Gamma(\frac{a_i}{2})\Gamma(\frac{a_j}{2})}{\Gamma(\frac{a_i+a_j}{2})}\ \ \ \ \ \ \ \ \ \ \ \ (i\not=j)\ .
\eea

In this way we reformulate  the original    problem as a problem of solving
of  the   ShG equation
\bea\label{luuausay}
\partial_w\partial_{\bar w}{\hat \eta}-\re^{2{ \hat \eta}}+ \re^{-2{ \hat \eta}}=0
\eea
in  the  open domain  ${\mathbb D}$ with the boundary 
\bea\label{saosisoai}
\partial  {\mathbb D}=\cup_{i=1}^3\{w_i\equiv {\bar w}_i\}\ .
\eea
The function 
\bea\label{apspsao}
{\hat \eta}={\tilde \eta}-{\textstyle \frac{1}{4}}\ \log(P{\bar P})= { \eta}-{\textstyle \frac{1}{4}}\ \log(p{\bar p})
\eea
should be real and smooth for all $z$ inside  ${\mathbb D}$ and 
possesses 
the following  asymptotic behavior 
near the boundary,
\bea\label{osasail}
{\hat\eta}=2\, l_i\ \log|w-w_i|+O(1)\ \ \ \ \ \ \ \ {\rm at}\ \ \ \ \ w\to w_i\ ,
\eea
with
\bea\label{sisiusaui}
l_i=\frac{2m_i+1}{a_i} -\frac{1}{2}\ \ \ \ \ \ \ \ (i=1,\, 2,\, 3)\ .
\eea


\subsection{Definition of the conserved charges}

The ShG equation \eqref{luuausay} possesses an infinite set of
continuity equations in the form
\bea\label{sksaklsalas}
\partial_{\bar w}{\hat F}_{2n}=\partial_w {\hat  G}_{2n-2}\ .
\eea
The functions $({\hat  F}_{2n},\,{\hat G} _{2n-2})$
are  conventional  tensor  densities.
They can be described  as follows. Let
\bea\label{sshjaqreew}
{\hat u}=(\partial_w{\hat \eta})^{2}-\partial^2_w{\hat \eta}\, ,\ \ \ \ \ \ \ \ \
{\hat v}=(\partial_w{\hat \eta})^{2}+\partial^2_w{\hat \eta}\ .
\eea
Then
\bea\label{ystksksksa}
{\hat F}_{2n} =U_{n}[\, {\hat u}\, ]\ ,\ \ \ \ \ \ \ \ {\hat G }_{2n-2}=
\re^{-2{\hat\eta}}\  U_{n-1}[\, {\hat v}\,]-
2\, \delta_{n,1}
\ ,
\eea
where $U_{n}[\,  {\hat u}\, ]$ are
homogeneous $({\rm grade}({\hat u})=2, \ {\rm grade}(\partial )=1,\
{\rm grade}(U_n)=2n )$
differential polynomials  in ${\hat u}$ of  degree
$n$  (known as the Gel'fand-Dikii polynomials \cite{Gelfand}),
\bea\label{iusksksak}
U_{n}[\,{\hat u}\,]=
\frac{2^n\,n!}{(2n-1)!!}\ \ {\hat \Lambda}^n\cdot 1\ .
\eea
Here
\bea\label{reqssksa}
{\hat \Lambda}=-{\textstyle \frac{1}{ 4}}\
\partial^2+ {\hat u}-{\textstyle \frac{1}{ 2}}\  {\partial}^{-1}\  {\hat u}'  \, ,
\eea
and prime stands for the derivative. Thus,
\bea\label{rwesksksa}
U_0[\, {\hat u}\,]&=& 1\ ,\nonumber\\
U_1[\, {\hat u} \,]&=&  {\hat u}\ ,\\
U_{2}[\,{\hat u} \,]&=& 
{\hat u}^2-{\textstyle \frac{1}{ 3}}\, {\hat u}''\ , \nonumber\\
U_{3}[\, {\hat u}\,]&=& 
{\hat u}^3-{\textstyle \frac{1}{2}}\, ( {\hat u} ')^2-
{\hat u}\, ''+
{\textstyle \frac{1}{ 5}}\ {\hat u}''''\ , \nonumber\\
U_{n}[\,{\hat u}\,]&=&
{\hat u}^n+\ldots\ ,\nonumber
\eea
where the last line shows overall normalization of the polynomials.
The continuity equations imply that
\bea\label{pospsaosp} 
\omega_{2n}=\re^{\ri\pi (n-\frac{1}{2})(a_1+a_2)}\, \big(\, \rd w\ {\hat F}_{2n}+
\rd {\bar w}\ {\hat  G}_{2n-2}\,\big)\
\eea
are closed one-forms.

Let us consider the integral
\bea\label{yustuissksksks}
{\mathfrak q}_{2n-1}=\int_{C_w}\omega_{2n}\ ,
\eea
where $C_w$ stands for some contour  in the open set  ${\mathbb D}$.
Generally speaking, the integral depends essentially on the choice of  integration contour.
However, for  a non-contractible loop
such that the local densities come to themself    when they are translated along $C_w$,
the integral 
is  not sensitive to continuous deformations  of $C_w$.
In this case ${\mathfrak q}_{2n-1}$
can be treated as a  conserved charge associated with
the closed one form $\omega_{2n}$.

In order to choose a suitable integration contour in
\eqref{yustuissksksks},
it is useful to make  the change of variables and return to the original
coordinates $z$ and ${\bar z}$.
Using the relation
\bea\label{isusaospsao}
\rd w=\re^{\frac{\ri\pi}{2} (a_1+a_2)}\ \sqrt{p(z)}\ \rd z\ ,
\eea
the integrand in   \eqref{yustuissksksks} can be rewritten in terms of
\bea\label{sopsaosa}
u=(\partial_z\eta)^2-\partial_z^2 \eta\ ,
\eea
which is a single-valued field on $\Sigma_{0,3}$.
For example, 
\bea\label{saosopsao}
{\hat u}= p^{-1}\ \left(  u+\frac{4 p p'' -5 p'^2}{ 16 p^2} \right)\ ,
\eea
and therefore, for $n=1$ the integral \eqref{yustuissksksks}
can written as
\bea\label{ksiauyalsklsa}
{\mathfrak q}_{1}= \int_{C} \bigg[\,\rd z\ \Big(\, \frac{u}{ \sqrt {p}} +\frac{1}{ 16 }\   p^{-\frac{5}{ 2}}\,  (4pp''-5p'^2)\,\Big)+
\rd {\bar z}\
{\sqrt{\bar p}}\ \ 2\, \big( \sqrt{p {\bar p}}\ \re^{-2 \eta}-1\big)\,\bigg]\ .
\eea
Here $C$ is an image of $C_w$ under the  inverse conformal transformation $w=w(z)$. 
Of course, it is straightforward  to perform the change of variables for any  given $n$.
We  do not present  explicit  formulas for $n>1$, just notice that
\bea\label{saospsao}
{\hat F}_{2n}[{\hat u}]=p^{-n}\ { F}_{2n}[u]+\ldots\ ,
\eea
where  the omitted terms contain derivatives of $u(z)$.

\begin{figure}
\centering
\includegraphics[width=16  cm]{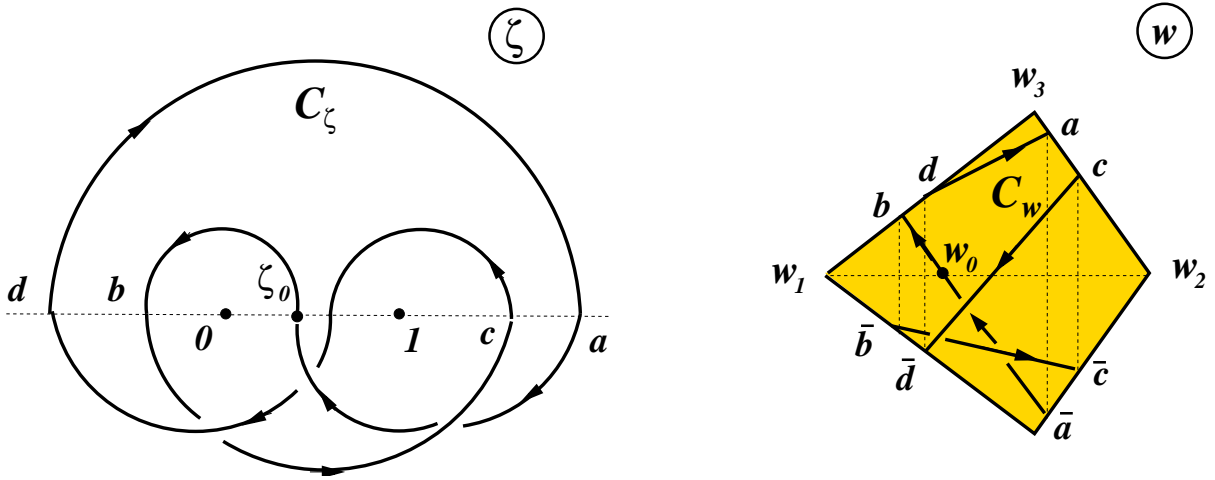}
\caption{The Pochhammer loop $C_\zeta$ with the base point $\zeta_0$. Its $w$-image
is homotopically equivalent to
a  union of the oriented  line segments
$C_w=
\protect\overrightarrow{ w_0 b}\cup
\protect\overrightarrow {{\bar b}{\bar c}}\cup
\protect\overrightarrow{ c {\bar d} }\cup  
\protect\overrightarrow{  d a}\cup
\protect\overrightarrow{ {\bar a} w_0}$.
The points $\{a,\,b,\,c,d\}$ in the $\zeta$-plane and their images are denoted by the same symbols.}
\label{fig2a}
\end{figure}

We now note that 
$\sqrt{p(z)}$ acquires  the same value
after  the analytic   continuation along the contour depicted in   Fig.\ref{fig1aa}.
Therefore the   integral  \eqref{ksiauyalsklsa} does not
depends on a base point   (an initial point of  the  integration path),  as well as the precise  shape of the 
contour. 
The image of this contour under 
the  M\"obius  transformation\ \eqref{sospsaos}
is the Pochhammer loop $C_\zeta$ shown in Fig.\,\ref{fig2a}. 
Recall that  the   Pochhammer loop, 
considered as 
an element of the fundamental group $\pi_1(\Sigma_{0,3})$,
is given by the commutator of the loops $\gamma_0$ and $\gamma_1$ which wind  anticlockwise  around  the 
points $\zeta=0$ and $\zeta=1$, respectively:
\bea\label{sapospsao}
C_\zeta= \gamma_0\circ\gamma_1\circ \gamma^{-1}_0\circ\gamma^{-1}_1\ .
\eea
Also, the complex conjugated contour ${\bar C}_\zeta$
is  homotopically equivalent  to the $C_\zeta$,
\bea\label{paopaos}
{\bar C}_\zeta=\gamma_{10}^{-1}\circ C_\zeta\circ \gamma_{10}\  \ \  \ \ \ {\rm with}\ \ \ \ \ \ \gamma_{10}=\gamma_1\circ\gamma_0\ .
\eea

This way we can  define the  conserved charges  by Eq.\eqref{yustuissksksks} where $C_w$ stands for the $w$-image (up to homotopy) of 
$C_\zeta$ (see Fig.\,\ref{fig2a}).
One can introduce another set of conserved charges
\bea\label{aoaspasopap}
{\bar {\mathfrak q}}_{2n-1}= \int_{{\bar  C}_w}{\bar \omega}_{2n}\ ,
\eea
where
\bea\label{aospsapsoap}
{\bar \omega}_{2n}=
\re^{\ri\pi (n-\frac{1}{2})(a_1+a_2)}\ \Big[\, \rd {\bar w}\ {\hat {\bar F}}_{2n}+
\rd { w}\ {\hat {\bar  G}}_{2n-2}\, \Big]\ ,
\eea
and the local densities  $ ({\hat {\bar F}}_{2n}, {\hat {\bar G}}_{2n-2})$ are  differential  polynomials
in ${\hat\eta}$ of the degree $2n$ which  are obtained from
 $ ({\hat {F}}_{2n}, {\hat {G}}_{2n-2})$ \eqref{ystksksksa}
by a formal substitution  $\partial_w\to \partial_{\bar w}$.
The integral is taken over the contour ${\bar C}_w$ which is complex conjugate to $C_w$.

Some explanation is in order here. Our definitions of the conserved charges
suffers from an intrinsic phase ambiguity of the form $\re^{\ri\pi n (N a_1+M a_2)}  \ (N,\,M\in {\mathbb Z})$ inherited from
the  phase ambiguity of  $\sqrt{p(z)}$.
In what follows 
we assume the condition
\bea\label{apospsposa}
\re^{\frac{\ri\pi}{2} (a_1+a_2)}\ \sqrt{p(z)}>0\ \ \ \ \ {\rm as}\ \ \ \ \ x_1<z<x_2<x_3\ ,
\eea
and define    $\sqrt{p(z)}$ outside of this real  domain by means of  the analytic continuation.
In terms of the coordinate $\zeta$ this convention implies that
the principal branch of  $\sqrt{P(\zeta)}$ in Eq.\eqref{soiosa}  is  chosen to satisfy   the condition
\bea\label{sopsosapas}
\sqrt{P(\zeta)}>0\ \ \ \ \  {\rm as}\ \ \ \ \ 0<\zeta<1\ .
\eea
For practical calculations, it is convenient to
rewrite  the  defining relations  \eqref{yustuissksksks},\,\eqref{aospsapsoap} in terms of the 
$(\zeta,{\bar \zeta})$-coordinates.
Then the integrals should be  taken along the
Pochhammer loops $C_\zeta$ and  ${\bar C}_\zeta$.  Bearing in mind   condition \eqref{sopsosapas},
it makes sense to chose the base point  $\zeta_0\in C_\zeta$ 
within the segment $[0,1]$ as  shown in  Fig.\,\ref{fig2a}. 
The
 values of the
integrands are determined unambiguously 
through  the analytic continuation along the integration path.
This convention resolves the issue of  phase ambiguity in the definitions of ${ {\mathfrak q}}_{2n-1}$ and ${\bar {\mathfrak q}}_{2n-1}$.
It  also implies that  the local densities  $ ({\hat {\bar F}}_{2n}, {\hat {\bar G}}_{2n-2})$   are complex conjugates
of $ ({\hat F}_{2n}, {\hat { G}}_{2n-2})$
\bea\label{aopspasosp}
{\hat {\bar F}}_{2n}={\hat F}_{2n}^*\ ,\ \ \ \ \ {\hat {\bar G}}_{2n-2}= {\hat { G}}_{2n-2}^*\ .
\eea
Now one can understand the r${\hat {\rm o}}$le of the phase factor $\re^{\ri\pi (n-\frac{1}{2})(a_1+a_2)}$
in the definitions of the conserved charges.
Indeed, one easily verifies  that Eqs.\eqref{paopaos} and \eqref{aopspasosp} yield  the complex conjugation condition
\bea\label{apospoapso}
{\bar {\mathfrak q}}_{2n-1}={ {\mathfrak q}}^*_{2n-1}\ .
\eea
In the case under consideration the  reflection symmetry $w\leftrightarrow{\bar w}$  of the
ShG equation is not broken
(i.e.  ${\hat\eta}(w,{\bar w})={\hat\eta}({\bar w},w)$), and therefore 
\bea\label{sospsoasaps}
{\bar {\mathfrak q}}_{2n-1}=
{\mathfrak q}_{2n-1}\ .
\eea
For this reason,  we can
focus on the integrals ${\mathfrak q}_{2n-1}$ only.

\subsection{\label{connection} A generating function for the conserved charges}

The  MShG equation
constitutes a flatness condition
for the  $sl(2)$ connection\footnote{
$\sigma^a$ are the usual Pauli matrices, i.e.,
$\sigma^3=
\begin{pmatrix}
1&0\\
0&-1
\end{pmatrix},\   \sigma^+=
\begin{pmatrix}
0&1\\
0&0
\end{pmatrix},\
\sigma^-=
\begin{pmatrix}
0&0\\
1&0
\end{pmatrix}$\ .}
\begin{eqnarray}\label{ssoiso}
{\boldsymbol A}_z&=&-
{\textstyle\frac{1}{ 2}}\ \partial_z\eta\ \sigma^3+\re^{\theta}\ \  \big[\,
\sigma^+\ \re^{\eta}+ \sigma^-\ p(z)\,\re^{-\eta}\, \big]\\
{ {\boldsymbol A}}_{\bar z} &=&
\ \ {\textstyle\frac{1}{2}}\ { \partial}_{\zb}\eta\ \sigma^3+
 \re^{-\theta}\, \big[\,
\sigma^-\ \re^{\eta}+ \sigma^+\ p({\bar z})\,\re^{-\eta}\, \big]\ ,
 \nonumber
\end{eqnarray}
with the spectral parameter $\theta$. 
The connection ${\boldsymbol A}={\boldsymbol A}_z\rd z+{\boldsymbol A}_{\bar z}\rd {\bar z}$ 
is not single-valued on the punctured sphere. However,
it does return to the original branch 
after the  continuation along the non-contractible loop
$C$ depicted in Fig.\,\ref{fig1aa}.
Therefore  the Wilson loop
\bea\label{sospsaosapo}
W(\theta)={\rm Tr}\bigg[{\cal P}\exp\Big(\oint_C {\boldsymbol  A}\,\Big)\,\bigg]
\eea
does not depend on the precise shape of the cycle used.

Apparently, $W(\theta)$ is 
an entire function of $\theta$. Since the shift of the argument $\theta\to\theta+\ri\pi$
can be compensated by the similarity transformation 
${\boldsymbol A}\to \sigma^3{\boldsymbol A}\sigma^3$, the Wilson loop is a periodic function of period $\ri\pi$
\bea\label{aiassaio}
W(\theta+\ri\pi)=W(\theta)\ .
\eea
It is easy to see that 
\bea\label{sospsoa}
W(\theta)\to 2\cosh\big( {\mathfrak q}_0\, \re^{\theta} +o(1)\,\big)\ \ \ \ {\rm as}\ \ \ \ 
\Re e(\theta)\to+\infty\,,\ \ |\Im  m(\theta)|<\frac{\pi}{2}\ ,
\eea
where 
\bea\label{osppapsopaos}
{\mathfrak q}_0=\oint_C\rd z\sqrt{p(z)}\ .
\eea
The integral  does not depend $z_i$ and
we can set them to the standard values $(0,1,\infty)$. Then the
integral over the Pochhammer loop $C_\zeta$  is performed using
well known
relation
\bea\label{opopsapospa}
\oint_{C_\zeta}\rd \zeta\ \zeta^{\alpha-1}(1-\zeta)^{\beta-1}=
\big(1-\re^{2\pi\ri\alpha}\big)\, \big(1-\re^{2\pi\ri\beta}\big)\ \frac{\Gamma(\alpha)\Gamma(\beta)}{\Gamma(\alpha+\beta)}\ ,
\eea
yielding the result
\bea\label{usysosppapsopaos}
{\mathfrak q}_0=-\frac{4\pi^2\rho\,\re^{\theta}}{\prod_{i=1}^3\Gamma(1-\frac{a_i}{2})}\ .
\eea
The subleading terms in the asymptotic expansion 
defined by the conserved charges.
The textbook calculation yields the following asymptotic expansions\ \cite{Faddeev:1987ph}: 
\bea\label{aaosioasao}
\log W(\theta)\sim
\begin{cases}
& - {\mathfrak q}_0\, \re^{\theta}+\sum_{n=1}^\infty c_n\  {\mathfrak q}_{2n-1}\ \re^{-(2n-1)\theta}\ \ \ 
{\rm as}\ \ \                            
\Re e(\theta)\to+\infty,\ \ |\Im  m(\theta)|<\frac{\pi}{2}\\
& - { {\mathfrak q}}_0\, \re^{-\theta}+\sum_{n=1}^\infty c_n\ {\bar {\mathfrak q}}_{2n-1}\ \re^{(2n-1)\theta}\ \ \
{\rm as}\ \ \ 
\Re e(\theta)\to-\infty,\ \ |\Im  m(\theta)|<\frac{\pi}{2}
\end{cases}
,
\eea
here $c_n=\frac{(-1)^{n}}{2 n!} \, \frac{\Gamma(n-\frac{1}{2})}{\sqrt{\pi}}$.
Thus $ W(\theta)$ can  be regarded as the generating function for the the conserved charges.

Notice that in the case of  unbroken reflection symmetry
the Wilson loop  is an even function of the spectral parameter
\bea\label{paopsao}
W(\theta)=W(-\theta)\ .
\eea


\section{Small $\rho$-expansion of the  first conserved charge}

Among the conserved charges,  $ {\mathfrak q}_1$ and ${\bar {\mathfrak q}_1}$
are of special interest.
Their linear combinations
\bea\label{sosipaosspaaps}
{\mathfrak h}={\mathfrak q}_{1}+{\bar {\mathfrak q}}_{1}\ ,\ \ \ 
\ \  \ \ \ \ \ \ {\mathfrak p}={\mathfrak q}_{1}-{\bar {\mathfrak q}}_{1}\ .
\eea
can be understood as the
energy and momentum of   the Pochhammer string. In the case under consideration,
the  momentum is zero.
Here we aim to explore the   small-$\rho$ expansion of ${\mathfrak q}_{1}$.

\subsection{On-shell value of the action functional}

\begin{figure}[!ht]
\centering
\psfrag{a}{ ${\mathbb  D}^{\rm (reg)}$}
\includegraphics[width=6  cm]{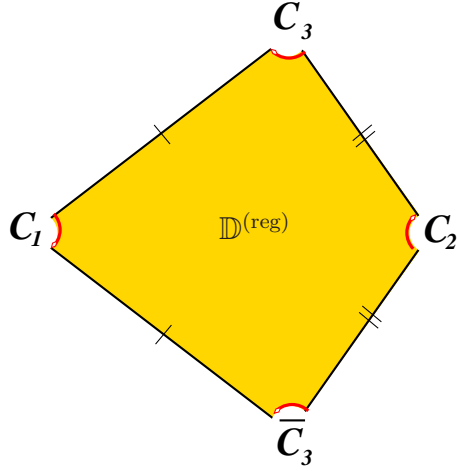}
\caption{The integration domain ${\mathbb  D}^{(\rm reg)}$ for the regularized action \eqref{saoisosais}. }
\label{fig9}
\end{figure}

The ShG  equation\ \eqref{luuausay} as well as the asymptotic conditions\ \eqref{osasail}  follow from the action
\bea\label{saoisosais}
{\cal A}[{\hat \eta}]=\lim_{\epsilon\to 0}\, \bigg[\, \frac{1}{\pi}\ \int_{{\mathbb D}^{\rm (reg)}}\rd^2 w\
\big(\, \partial_w{\hat\eta}\partial_{\bar w}{\hat\eta}+4\,\sin^2({\hat\eta})\, \big)+
\sum_{i=1}^3\frac{l_i}{\pi\epsilon}\ \oint_{C_i}{\rd \ell}\, {\hat \eta}-\log(\epsilon)\ 
\sum_{i=1}^3 a_i\,l_i^2\,\bigg]\ .
\eea
The  first term here  contains a  two-fold  integral\footnote{In this work
we always  use the shortcut notation for the 2D integration measure:  ${\rd^2 w}:=\frac{1}{2\ri}\ \rd w\wedge\rd{\bar w}$.}
over the domain
\bea\label{sksosa}
{\mathbb D}^{\rm (reg)}={\mathbb  D}\backslash \cup_{i=1}^3{\mathbb  D}^{(i)}_\epsilon\, ,\ \ \ {\mathbb  D}^{(i)}_\epsilon=
\{ w\in {\mathbb  D}\ :\ 
| w-w_i|<\epsilon\}
\eea
depicted in Fig.\,\ref{fig9}.
It  is the ``cutoff'' version of the naive action functional.
The additional terms involve  the  integrals  over  the  boundary  $\partial {\mathbb  D}^{\rm (reg)}$
(a union of the
three closed  contours    $C_{1}$, $C_{2}$ and $C_3\cup{\bar C}_3$) 
and field-independent ``counterterms'' which provide  an existence of the limit.
We define  the on-shell action ${\cal A}^*$ as
a value of the functional ${\cal A}[{\hat \eta}]$
calculated on a solution  of  the problem \eqref{luuausay},\,\eqref{osasail}.

It was already mentioned that 
the overall size of the polygon 
is   defined  by the homothety parameter $\rho$.
As $\rho \to+\infty$
the dominating contributions to  the on-shell action 
come from the boundary $\partial {\mathbb D}$.
Near the vertexes, ${\hat \eta}$ can be expressed in terms of 
certain  Painlev${\acute{\rm e}}$ III transcendents. This   observation allows one  to determine (see, e.g., 
Appendix B from Ref.\cite{Lukyanov:2011wd})
the limiting value of the on-shell action
\bea\label{sisoisai}
{\cal A}^*_\infty=\lim_{\rho \to+\infty}{\cal A}^*\ .
\eea
The result reads as follows
\bea\label{apososa}
{\cal A}^*_\infty=
-\sum_{i=1}^3a_i\  \Big(\, l^2_i\ \log(2)+F\big(l_i+{\textstyle\frac{1}{2}}\big)\, \Big)\ ,
\eea
where we use the notations
\bea\label{asosapsa}
F(x)=\int_{\frac{1}{2}}^x\rd t\, \log\gamma(x)\ ,\ \ \ \ \ \ \gamma(x)=\frac{\Gamma(x)}{\Gamma(1-x)}\ .
\eea
It makes sense to subtract the constant\ \eqref{sisoisai} from the action\ \eqref{saoisosais} and define 
\bea\label{apsoisaoi}
{\mathfrak Y}={\cal A}^*-{\cal A}^*_\infty\ .
\eea
This function is  the on-shell action  normalized by the condition
\bea\label{opapsasops}
\lim_{\rho\to+\infty}{\mathfrak Y}=0\ .
\eea
We shall refer to  ${\mathfrak Y}={\mathfrak Y}(\rho|\{l_i,\,\alpha_i\})$ as to the ``Yang-Yang function'' 
(for general explanation of this term, 
see Ref.\cite{Nekrasov:2009rc}. In this work 
we closely follow the notations from Ref.\cite{Lukyanov:2011wd}).

\subsection{Small-$\rho$ expansion of the Yang-Yang function}

To explore the small $\rho$-limit it is useful to return to the original
complex  coordinates $(z,{\bar z})$ and the field $\eta$.
The action  \eqref{saoisosais} can be rewritten in the form (see Appendix A for details):
\bea\label{tystsaooiausais}
{\cal A}[ \eta]&=&{\cal A}_0+\lim_{\epsilon_i\to 0\atop
R\to \infty}
\, \bigg[\, \frac{1}{\pi}\ \int_{|z-z_i|>\epsilon\atop
|z|<R}\rd^2 z\ 
\big(\, \partial_z{\eta}\partial_{\bar z}{\eta}+\re^{2\eta}+p(z){\bar p}(z)\,\re^{-2\eta}\,\big)
\\
&+&2\ \sum_{i=1}^3 \big(\,m_i\,\eta_i- m_i^2\,\log(\epsilon_i)\,\big)
+2\,\eta_\infty+2\,\log R\, \bigg]\ .\nonumber
\eea
Here
\bea\label{tsrapsopsa}
 \eta_i=\frac{1}{2\pi\epsilon_i}\ \oint_{|z-z_i|=\epsilon_i} \rd \ell \, \eta\ ,\ \ \ \ \ \
\eta_\infty=\frac{1}{2\pi R}\ \oint_{|z|=R}\rd \ell\,\eta\ ,
\eea
whereas an  explicit form of the field-independent constant ${\cal A}_0$ is given by 
the equation \eqref{sospsaapsopa} from  Appendix A.

In the vicinity of   punctures one has
\bea\label{saosoas}
\re^{2\eta}\propto |z-z_i|^{4 m_i}\ ,\ \ \ \ \ \ \ \ 
p(z){\bar p}({\bar z})\ \re^{-2\eta}\propto  \rho^4\ |z-z_i|^{2(a_i-2 m_i-2)}\ .
\eea
Therefore,
assuming that
\bea\label{isaosaoias}
-{\textstyle \frac{1}{2}}<m_i<  -{\textstyle \frac{1}{2}}+{\textstyle \frac{1}{4}}\ a_i\ ,
\eea
we may  ignore the  term $p(z){\bar p}({\bar z})\,\re^{-2\eta}$  in the functional \eqref{tystsaooiausais} as  $\rho \to 0$.
In other words, the small-$\rho$ limit is controlled by  the classical  Liouville theory on the sphere with
three punctures, and the following  limit does exist for any $z\not= z_i$: 
\bea\label{sospsaopsao}
\lim_{\rho \to 0}\eta(z,{\bar z})= {\textstyle \frac{1}{2}}\ \varphi(z,{\bar z})\ .
\eea
Here the field $ \varphi$
is a solution of the Liouville equation
\bea\label{opssopsaas}
\partial_z\partial_{\bar z}\varphi=2\ \re^{\varphi}\ ,
\eea
subject of  the 
boundary conditions
\bea\label{asposapas}
\varphi(z,{\bar z})&=&4\, m_i\,\log|z-z_i|+O(1)\ \ \ \ \ \ {\rm at}\  \ \ z\to z_i\nonumber\\
\varphi(z,{\bar z})&=&-4\, \log|z|+O(1)\ \ \ \ \ \ \ \ \ \ \ \ \ \ {\rm at}\ \ \ z\to \infty\ .
\eea
This is exactly the problem of  constructing  a  metric of  constant intrinsic curvature
\bea\label{sskjsajks}
{\rd s}^2_{\rm cic}=\re^{\varphi}\ \rd z \rd {\bar z} 
\eea
on the sphere
with three punctures.
The total intrinsic curvature is given by
\bea\label{saospsasoa}
\int_{\Sigma_{0,3}}{\tt K}[\varphi]=4\pi\ \Big(1+\sum_{i=1}^3 m_i\Big)\ .
\eea
Since $m_i$ obey inequality
\eqref{isaosaoias}, then
$-2\pi < \int_{\Sigma_{0,3}}{\tt K}<0$.  Therefore
the solution
exists and defines
a metric  of  constant negative  Gaussian curvature on $\Sigma_{0,3}$.
The corresponding on-shell value  of the Liouville regularized action
was calculated  in Ref.\cite{Zamolodchikov:1995aa}.
Using this result it is straightforward
to derive the small-$\rho$ behavior of
the Yang-Yang function \eqref{apsoisaoi} (see also the end of Appendix A for some technical details):
\bea\label{ystasosposa}
{\mathfrak Y}=\log(\rho)\ \sum_{i=1}^3\Big(\frac{1}{6}-\frac{4 }{ a_i}\ p_i^2\, \Big)+
{\mathfrak Y}_0-2\, \rho^2\, \prod_{i=1}^3\gamma\Big({\frac{a_i}{2}}\Big)+
o(\rho^2)\ .
\eea
Here
\bea\label{sopssapis}
p_i=m_i+{\textstyle \frac{1}{2}}
\eea
and the $\rho $-independent constant 
is expressed in terms of the function $F$ \eqref{asosapsa},
\bea\label{sossaiasi}
&&{\mathfrak Y}_0=
F\big({\textstyle\frac{1}{2}}-p_1-p_2-p_3\big)+
F\big({\textstyle\frac{1}{2}}-p_1-p_2+p_3\big)+
F\big({\textstyle\frac{1}{2}}-p_2-p_3+p_1)\\
&&+F\big({\textstyle\frac{1}{2}}-p_3-p_1+p_2\big)-F(0)
+\sum_{i=1}^3\Big(\, a_i\, F\big({\textstyle\frac{2 p_i}{a_i}}\big)-
F(1-2p_i)+a_i\, \big({\textstyle\frac{2p_i}{a_i}}-
{\textstyle\frac{1}{2}}\big)^2\ \log(2)\,\Big)\ .\nonumber
\eea

Eq.\eqref{ystasosposa} shows the nature of singular behavior at $\rho \to 0$. 
Notice that 
the term $p(z){\bar p}({\bar z})\,\re^{-2\eta}$ in the action \eqref{tystsaooiausais} can be treated as 
a uniformly bounded perturbation $\propto \rho^4$,
as far as the condition\ \eqref{isaosaoias} is fulfilled.\footnote{
It seems likely that the formal proof of existence and uniqueness solution of 
the problem \eqref{akasjksa},\,\eqref{asosospa} can be obtained basing on this observation.}
Thus,
the Yang-Yang function \eqref{apsoisaoi}
admits the  small-$\rho$ expansion
of the form
\bea\label{asosposa}
{\mathfrak Y}=\log(\rho)\ \sum_{i=1}^3\Big(\frac{1}{6}-\frac{4 }{ a_i}\ p_i^2\, \Big)+
{\mathfrak Y}_0-
2\, \rho^2\, \prod_{i=1}^3\gamma\Big({\frac{a_i}{2}}\Big)
+\sum_{n=1}^\infty y_n\  \rho^{4n}\ .
\eea

\subsection{
\label{infdel}Infinitesimal homothety of the  flat metric of punctured sphere}

The variation of the action under  infinitesimal perturbations of the 
moduli  of the  flat metric on $\Sigma_{0,3}$ do not vanish on-shell.
They can be expressed in terms of  the stress-energy tensor.
Here we focus on  the infinitesimal homothety.

For an   infinitesimal dilation
$\frac{\delta \rho}{\rho}=\frac{\delta\epsilon}{\epsilon}=\lambda \ll 1$, one has
\bea\label{xsososa}
\delta_\rho  {\mathfrak Y}=\frac{\delta \rho}{\rho} \, \bigg[\,
\lim_{\epsilon\to 0}\, \frac{2}{\pi}\int_{D^{\rm (reg)}} \rd^2w\  \Theta-
\sum_{i=1}^3 a_i\,l_i^2\, \bigg]\ ,
\eea
where $\Theta$ stands for   the  trace of
the stress-energy tensor of the  ShG equation,
\bea\label{apospaosp}
\Theta=4\,\sinh^2({\hat\eta})\ .
\eea
The calculation outlined in Appendix B yields the formula which expresses
the difference in the bracket 
in   Eq.\eqref{xsososa}  in terms of the
conserved charge  ${\mathfrak q}_1={\bar {\mathfrak q}}_1$ and
the circumdiameter $d$ of the triangle   $ (w_1,w_2,w_3)$ in  Fig.\,\ref{fig8}:
\bea\label{aposussay}
\rho\  \frac{\partial  {\mathfrak Y}}{\partial \rho}=- {\mathfrak F}\ ,
\eea
with
\bea\label{sissisusaopsopsaps}
{\mathfrak F}=\frac{d}{4\pi}\ ({\mathfrak q}_1+{\bar {\mathfrak q}_1})\ ,\ \ \ \
d=\frac{\rho}{\pi}\ \prod_{i=1}^3\Gamma\Big(\frac{a_i}{2}\Big)\ .
\eea
Eq.\eqref{opapsasops} implies that
${\mathfrak F}$ is normalized by the condition
\bea\label{ystasopspsao}
\lim_{\rho\to+\infty}{\mathfrak F}=0\ .
\eea
It is easy to see that  ${\mathfrak F}$ can be represented in a form
\bea\label{fasosposa}
{\mathfrak F}=-\frac{1}{6}\,\sum_{i=1}^3\Big(1-\frac{24 }{ a_i}\ p_i^2\, \Big)+
4\, \rho^2\, \prod_{i=1}^3
\gamma\Big({\frac{a_i}{2}}\Big)-\frac{4}{\pi}\ \int_{\Sigma_{0,3}} \rd^2 z\ p(z){\bar p}({\bar z})\, \re^{-2\eta}\ ,
\eea
and the small-$\rho$ expansion  \eqref{asosposa} yields\footnote{Notice that, as it follows from the  MShG equation,
$$\frac{1}{\pi}\ \int_{\Sigma_{0,3}} \rd^2 z\  \re^{2\eta}=
\frac{1}{\pi}\ \int_{\Sigma_{0,3}} \rd^2 z\ p(z){\bar p}({\bar z})\, \re^{-2\eta}
-\frac{1}{2}\,\sum_{i=1}^3 a_i l_i\ . $$}
\bea\label{aspospas}
\frac{1}{\pi}\ \int_{\Sigma_{0,3}} \rd^2 z\ p(z){\bar p}({\bar z})\, \re^{-2\eta}
=
\sum_{n=1}^\infty n\,y_n\  \rho^{4n}\ .
\eea
The first coefficient in this series
is
simply expressed
in terms of the Liouville field $\varphi$\ \eqref{sospsaopsao},
\bea\label{ystsapossposai}
y_1=\frac{1}{\pi \rho^4}\ \int
\rd^2 z\
p(z)\,{\bar p}({\bar z})\ \re^{-\varphi(z,{\bar z})}\ .
\eea
An explicit  analytical expression  for $\varphi$
can be found in Ref.\cite{Zamolodchikov:1995aa} (see Eqs.(4.1)-(4.5) therein).

\section{\label{qn}Small-$\rho$ limit of  ${\mathfrak q}_{2n-1}$ for $n>1$}

Eqs.\eqref{sissisusaopsopsaps}-\eqref{aspospas} imply that
\bea\label{sopsopsaspaa}
\lim_{\rho\to 0}\,(\rho\, {\mathfrak q}_1)=\frac{8\pi^2}{\prod_{i=1}^3\Gamma(\frac{a_i}{2})}\ 
\bigg(\,\sum_{i=1}^3\frac{p_i^2}{a_i}-\frac{1}{8}\, \bigg)\ .
\eea
It is instructive to verify this result using  the original definition \eqref{ksiauyalsklsa}.

The limiting value of the field $\eta$ is expressed in terms of the solution of
the Liouville equation. Therefore  as $\rho\to 0$
the field $u(z,{\bar z})$ turns to be a holomorphic component of the  Liouville stress-energy tensor:
\bea\label{saossaposa}
\lim_{\rho\to 0} u(z,{\bar z})=T_L(z)\ \ \ \ \ {\rm with}\ \ \ \ \ T_L(z)=\frac{1}{4}\ (\partial_z\varphi)^2-\frac{1}{2}\ \partial^2_z\varphi\ . 
\eea
In the case of the sphere with three punctures\ \eqref{asposapas}
the holomorphic fields $T_L(z)$ has the form 
\bea\label{aspospospas}
T_L(z)=-\sum_{i=1}^3\bigg(\, \frac{\delta_i}{(z-z_i)^2}+\frac{c_i}{z-z_i}\,\bigg)\ ,\ \ \ \  \ \ {\rm where}\ \ \ 
\delta_i=\frac{1}{4}-p_i^2\ .
\eea 
Since $z=\infty$ is a regular point,
$T_L(z)\propto \frac{1}{z^4}$ as $ z\to \infty$. This imposes three linear relations on $c_i$
which  determine them uniquely,
\bea\label{sopspsosa}
c_i=\frac{\delta_i+\delta_j-\delta_k}{z_j-z_i}+\frac{\delta_i+\delta_k-\delta_j}{z_k-z_i}\ ,\ \  \ \ \ (i,j,k)={\tt perm}(1,2,3)\ .
\eea
Combining  Eqs.\eqref{ksiauyalsklsa} and \eqref{saossaposa}, one gets
\bea\label{ksiauyalsklsausys}
{\mathfrak q}_{1}=\oint_{C} \frac{\rd z}{\sqrt {p(z)}}\left[\    T_L(z) +
\frac{1}{ 4 }\  \frac{\partial_z^2p(z)}{p(z)}-\frac{5}{16} \ 
\left(\frac{\partial_z p(z)}{p(z)}\right)^2\ \right]+O(1)\  \ \ \ \ \ \ {\rm as}\ \ \ \rho \to 0\ ,
\eea
where the contour is shown in Fig.\,\ref{fig1aa}. Let  $z_1<z_2<z_3$, then  the base point of  $C$  
should be taken inside  the interval $[ z_1, z_2]$, whereas  the branch of the multivalued function $\sqrt{p(z)}$ is
determined by the condition \eqref{apospsposa}. 
With this convention,  the integral does not depend on $z_i$ and
we can set them to be  $(0,1,\infty)$. Then using  relation \eqref{opopsapospa}, 
one easily reproduces Eq.\eqref{sopsopsaspaa}.

For $n>1$, the leading small-$\rho$   behavior of ${\mathfrak q}_{2n-1}$
can be obtained similarly.  The calculations are elementary but  rather cumbersome.
For example, for  $n=2$, the result can be written in the form
\bea\label{sjusyu}
\lim_{\rho\to 0}(\,\rho^{3}\, {\mathfrak q}_{3}\,) &=&-\frac{2\pi^2}{3\,\prod_{i=1}^3\Gamma(\frac{3a_i}{2})}\ \bigg[\,
 \sum_{i=1}^3E_i\, \Big(\frac{x_i^2}{16}-\frac{x_i}{ 16}+\frac{1}{ 192}\Big)\\
&+&
\sum_{
i\not= j}E_{ij}\, \Big(\frac{x_i}{4}-\frac{1}{ 24}\Big)
\Big(\frac{x_j}{4}-\frac{1}{ 24}\Big)+
\frac{1}{ 240}\
\sum_{i=1}^3H_{i}\ \,\bigg]\ \ \ \ \ \ \ \Big(\,x_i=\frac{4p^2_i}{a_i}\,\Big)\ ,
\nonumber
\eea
where the numerical coefficients $E_i,\ E_{ij}$ and $H_i$ are given by
\bea\label{kasuaiu}
&&E_{i}=a_i\, (3a_j-2)\,(3a_k-2)
\nonumber\\
&&E_{ij}=3\, a_i\, a_j\, (3a_k-2)
\\
&&H_i=8-a_i^2-9\ (a_1a_2+a_2a_3+
a_3a_1)+
15\ a_1a_2a_3
\ .\nonumber
\eea
In these equations $(i,\,j,\,k)$ represents any permutation of the numbers
$(1,2,3)$.
It is much easy to  establish the following general structure,
\bea\label{aopspaaspo}
\lim_{\rho\to 0}(\,\rho^{2n-1}\, {\mathfrak q}_{2n-1}\,)=R_{n}
\bigg(\,\frac{4p^2_1}{a_1},\,\frac{4p^2_2}{a_2},\,\frac{4p^2_3}{a_3}\,\bigg)
\ ,
\eea
where  $R_{n}$ stand   for  polynomials in  the variables  $x_i=\frac{4p_i^2}{a_i}$  of degree $n$,
\bea\label{spospospa}
R_{n}(x_1,x_2,x_3)=
\sum_{i+j+k=n} R^{(n)}_{ijk}\ \ x^i\,x^j\,x^k
+\ldots\ .
\eea
The dots here  represent   the  sum of monomials of   degree lower than $n$.
It is not difficult to calculate    the highest coefficients for any $n$,
\bea\label{usisusspospospa}
&&R^{(n)}_{ijk}=
\frac{(-1)^{n-1}\ 2^{5-2n}  \pi^2}{\prod_{i=1}^3\Gamma((n-\frac{1}{2})\,a_i)}\ \ 
\frac{ n!\, \big(a_1(\frac{1}{2}- n)\big)_{n-i} \big(a_2\,(\frac{1}{2}- n)\big)_{n-j}\,\big(a_3\,(\frac{1}{2}- n)\big)_{n-k}}
{ i!\,j!\,k!\ \, (2n-1)^3\  a^{1-i}_1a^{1-j}_2a^{1-k}_3}\ .
\eea

\section{\label{id}Identification of the parameters}

We are now in position to relate the parameters of the problem \eqref{akasjksa},\,\eqref{asosospa} and 
the couplings of   the Lagrangian \eqref{aposoasio}.
As  it was mentioned, the Fateev model has infinitely many local integrals of motion.
The displayed terms in Eq.\eqref{isusospsasopas} fix the normalization of these operators.
Let  $I_{2n-1}=I^{(+)}_{2n-1}=I^{(-)}_{2n-1}$  be the vacuum eigenvalues of  ${\mathbb I}^{(+)}_{2n-1}$ and
${\mathbb I}^{(-)}_{2n-1}$.
In the CFT limit, i.e. at  $\mu=0$ in  the Lagrangian \eqref{aposoasio},
these functions become polynomials in $k_i$ of the degree $n$,\footnote{
In this limit, \eqref{sapsapo} acquires continuous symmetries with respect to any shifts of the fields
$\varphi_i$,  and the limiting values \eqref{ustsrslslssaystsr} are  no longer periodic in $k_i$.} and 
the normalization in \eqref{isusospsasopas} is such that
at $\mu=0$ we have
\bea\label{ustsrslslssaystsr}
I_{2n-1}|_{\mu=0}=
\Big(\frac{2\pi}{R}\Big)^{2n-1}\ \bigg[
\sum_{i+j+k=n} C^{(n)}_{ijk} \  \big(2\alpha_1 k_1\big)^{2 i}\ \big(2\alpha_2 k_2\big)^{2j}\, \big(2\alpha_3 k_3\big)^{2k}
+\ldots\, \bigg]\, .
\eea
The polynomials are known  explicitly for  $n=1,\,2$,  whereas the constant 
$ C^{(n)}_{ijk}$ is known for any $n$ and given by Eq.\eqref{saosopsaosa}.
All these  results were obtained in Ref.\cite{Lukyanov:2012wq}.
In a view of Eqs.\eqref{sopsopsaspaa},\,\eqref{sjusyu}-\eqref{usisusspospospa},
they are  all in agreement with the proposal \eqref{aspspsapo}, provided  
that the    identification \eqref{sssaopsa} and the relation involving the normalization constant $d_n$,
\bea\label{osposapa}
d_{n}\  \ \ \frac{(-1)^{n-1}\  16 \pi^2}{\prod_{i=1}^3\Gamma(\,2\,(2n-1)\,\alpha^2_i\,)}\ \ 
\lim_{\mu R\to 0}\Big(\frac{\mu R}{4\pi\rho }\Big)^{2n-1}=1\ ,
\eea
are imposed. 

To find the relation between $\rho$ and  the dimensionless parameter $\mu R$,
let us consider the
small-$R$   expansion  of the finite-volume energy \eqref{apospsapo}.
A brief inspection of the Lagrangian  \eqref{aposoasio}  reveals
that
\bea\label{asopisspaasop}
\frac{RE}{\pi}=-\frac{c_{\rm eff}}{6}-\sum_{n=1}^\infty e_n\ (\mu R)^{4n}\ .
\eea 
Here the first term is dictated  by the simple Gaussian model
underlining  the CFT  limit  with    the   effective central charge
\bea\label{posapsapoas}
c_{\rm eff}=\sum_{i=1}^3\big(1-24\,\alpha^2_i\,k_i^2\,\big)\ .
\eea
The general structure of the 
short distance  expansion follows from the fact that the potential term in the Lagrangian \eqref{aposoasio}
is a uniformly bounded perturbation for any finite value of the dimensionless parameter $\mu R$. Therefore
the Conformal Perturbation Theory can be  applied literally and
yields an expansion of the form \eqref{asopisspaasop} with
coefficients $e_n$ are  expressed in terms of certain $2D$ Coulomb-type integrals.
The large-$R$ behavior of \eqref{asopisspaasop} is defined by the specific  bulk energy \eqref{saospsaos},\,\eqref{asososa}.
Eqs.\eqref{ystasopspsao}-\eqref{aspospas} strongly suggest the following  identification
\bea\label{isuaosposao}
{\mathfrak F}=\frac{R}{\pi}\ \big(\,E-R\,{\cal E}_0\,\big)\ .
\eea
This is, in fact, the first line in \eqref{aspspsapo}, provided that $\frac{1}{2}\ \mu R$ coincides with $\rho$ and
the coefficients $e_n$  in  expansion  \eqref{asopisspaasop} are simply   related to
the coefficients  $y_n$ from the
small-$\rho$ expansion of the Yang-Yang function\ \eqref{ystasosposa},\footnote{
For $n=1$, Eq.\,\eqref{asosapaso}  combined  with \eqref{ystsapossposai}
leads to a non-trivial prediction for the coefficient   $e_1$. It would be interesting
to confirm     this  result  within  the Conformal Perturbation Theory.}
\bea\label{asosapaso}
e_n= 2^{2-4n}\ n\,y_n\ .
\eea
Finally, Eq.\eqref{osposapa} yields the formula \eqref{poapsospsaos} for the coefficient $d_n$.

The  $\mu-\rho$ relation \eqref{saossaops} implies
that the bulk specific energy  is simply expressed  in terms of
the area $A$ of  the punctured sphere $\Sigma_{0,3}$ calculated
w.r.t. the flat metric \eqref{oiaoisao} (see Eq.\eqref{rssaosiosa} in Appendix A):
\bea\label{rssjsjsusaosiosa}
R^2\,{\cal E}_0=-4\, A\ .
\eea
In order to explain  the meaning
of the 
lengths of  the sides $|w_i-w_j|$ \eqref{soosapsosa} in the Fateev model, let us recall some
facts concerning  the factorizable  scattering  theory associated with this QFT. All the details can be found in Appendix F  in
Ref.\cite{Fateev:2004un}. 

The spectrum consists of three quadruplets of fundamental particles 
\bea\label{asosopsapsa}
Z_{\epsilon\epsilon'}^{(i)}\ , \ \ \ \ \ \ \ \ \ \epsilon,\,\epsilon'=\pm\,,\ \ \ \  i=1,\,2,\,3\ ,
\eea
with the masses
\bea\label{apsaisaosao}
M_i=M_0\ \sin\Big(\frac{\pi a_i}{2}\Big)\ ,\ \ \ \ M_0= \frac{2\mu}{\pi}\ \prod_{i=1}^3\Gamma\Big(\frac{a_i}{2}\Big)\ 
\eea
and their bound states. (Here the relation  $a_i=4\alpha_i^2$ is assumed to hold.) 
The Zamolodchikov-Faddeev commutation  relations for the fundamental particles
read
\bea\label{sopspospas}
&&Z_{\epsilon_1\epsilon'_1}^{(i)}(\theta_1)Z_{\epsilon_2\epsilon'_2}^{(i)}(\theta_2)=
-\sum_{\epsilon_3\,\epsilon'_3\atop
\epsilon_4\,\epsilon'_4}
\big[\,S_{a_j}(\theta_1-\theta_2)\,\big]^{\epsilon_3\epsilon_4}_{\epsilon_1\epsilon_2}\
\big[\,S_{a_k}(\theta_1-\theta_2)\,\big]^{\epsilon'_3\epsilon'_4}_{\epsilon'_1\epsilon'_2}\
Z_{\epsilon_4\epsilon'_4}^{(i)}(\theta_2)Z_{\epsilon_3\epsilon'_3}^{(i)}(\theta_1)\nonumber\\
&&Z_{\epsilon\epsilon'_1}^{(i)}(\theta_1)Z_{\epsilon'_2\epsilon''}^{(j)}(\theta_2)=\epsilon\,\epsilon''
\sum_{\epsilon_3\,\epsilon'_4}
\big[\,{\hat S}_{a_k}(\theta_1-\theta_2)\,\big]^{\epsilon'_3\epsilon'_4}_{\epsilon'_1\epsilon'_2}\
Z_{\epsilon_4\epsilon''}^{(j)}(\theta_2)Z_{\epsilon\epsilon'_3}^{(i)}(\theta_1)\ ,
\eea
where $(i,j,k)={\tt perm}(1,2,3)$ and
\bea\label{aspopsao}
{\hat S}_{a}(\theta)=\ri\ \tanh\big({\textstyle\frac{\theta}{2}}+
\ri\, {\textstyle\frac{\pi a}{2}}\,\big)\ S_a\big({\textstyle\frac{\theta}{2}}+
\ri\, {\textstyle\frac{\pi a}{2}}\,\big)\ .
\eea
Also $S_a(\theta)$ stands for the conventional $S$-matrix in the quantum sine-Gordon theory \cite{Zamolodchikov:1978xm}
with the renormalized coupling constant  $a$, 
related to  the Coleman coupling $\beta^2_{C}$ \cite{Coleman:1974bu}
as follows
\bea\label{sosposa}
a=\frac{\beta^2_C}{8\pi-\beta_C^2}\ .
\eea
 
It is easy to see that 
the $\mu-\rho$ relation  implies that the  dimensionless  parameter 
$\frac{1}{4}\ RM_0$ is the circumdiameter of the triangle $w_1 w_2 w_3$ from Fig.\,\ref{fig8},
whereas  $\frac{1}{4}\, R\,M_i$ are   given by the 
lengths of  the corresponding sides:
\bea\label{saosaposa}
R M_i=4\ |w_j-w_k|\ ,\ \ \  \ \ {\rm where}\ \ \ \ \ \  (i,j,k)={\tt perm}(1,2,3) \ .
\eea

\section{Concluding remarks}

This work did not aim to achieve  rigorous derivation of the ODE/IM correspondence.
The goal was 
to propose a  meaningful relation
between  a  certain   problem for the    MShG equation
on the one hand, and   the  Fateev model on the other.
Forwarding  can be performed along the following line.

Having at hand Eqs.\eqref{sopspospas}, 
it is not hard to
guess the  Non Linear    Integral Equations (NLIE) \cite{Klumper:1991,Destri:1992qk}  
associated  with this factorizable scattering theory.
In fact, the system of NLIE
was   already 
proposed
 by  Fateev in his unpublished work \cite{FatevU}.
Unfortunately, it still lacks the first principle derivation, say, from the
lattice.\footnote{In the limiting case $\alpha_3^2=0$ (the so called Bukhvostov-Lipatov model)
the  NLIE were  derived from the coordinate Bethe Ansatz in Ref.\cite{Saleur:1998wa}.}
Nevertheless it would be important to confirm   Fateev's NLIE from the ODE side.
The part of the work was already done by  Bazhanov.

In order to explain the main  result of the unpublished  paper \cite{BazU}, 
let us  consider the auxiliary problem \eqref{lp},\,\eqref{ssoiso}.
As is well  known, this  matrix system
can be reduced to second order linear
differential equations. One can write general solution of
\eqref{lp} as
\bea\label{ilslsa}
{\boldsymbol \Psi}=
\begin{pmatrix}
\re^{\frac{\theta}{ 2}}\ \re^{\frac{\eta}{ 2}}\ \psi\\
\re^{-\frac{\eta}{ 2}}\ \re^{-\frac{\theta}{
2}}\,(\partial_z+\partial_z\eta)\, \psi
\end{pmatrix}=
\begin{pmatrix}
\re^{-\frac{\eta}{ 2}}\  \re^{\frac{\theta}{ 2}}\, ( \partial_{\zb}+
\partial_{\zb} \eta)\,   {\bar \psi}\\
\re^{\frac{\eta}{ 2}}\ \re^{-\frac{\theta}{ 2}}\ {\bar\psi}
\end{pmatrix}\ ,
\eea
where $\psi$ and ${\bar \psi}$ solve the equations
\bea\label{sksaksayst}
&&\big[\, \partial_{z}^2-u(z,{\bar z})-\re^{2\theta}\ \ p(z)\, \big]\ \psi=0\\
&&
\big[\, \partial^2_{\zb}-{\bar u}(z,{\bar z})-\re^{-2\theta}\,
{ p}({\bar z})\, \big]\ {\bar \psi}=0\ .\nonumber
\eea
Let us focus on the first equation in \eqref{sksaksayst}.
This form is convenient for taking  the limit $\rho\to 0$,  
when the field $u(z,{\bar z})$ 
turns to be the  holomorphic component of the  Liouville stress-energy  tensor $T_L(z)$ \eqref{aspospospas}.
The function $p(z)$\ \eqref{sossao} contains the  small parameter $\rho^2$ as an  overall factor, which
can be absorbed by a suitable
shift of the spectral parameter. Thus
we define  the new  parameter
\bea\label{saopsoaa}
\kappa=\rho\ \re^\theta
\eea
and will keep it  fixed  as $\rho\to 0$. This yields
the ODE
\bea\label{sksaksa}
\big[\, -\partial_{z}^2+V_0(z)+V_1(z)\, \big]\ \psi=0\ ,\
\eea
with
\bea\label{asopaopaosp}
V_0(z)&=&
-\sum_{i=1}^3\bigg(\, \frac{\delta_i}{(z-z_i)^2}+\frac{c_i}{z-z_i}\,\bigg)
\nonumber
\\
V_1(z)&=&\kappa^2\ \frac{(z_3-z_2)^{a_1}\,(z_1-z_3)^{a_2}\,(z_2-z_1)^{a_3}}
{(z-z_1)^{2-a_1}(z-z_2)^{2-a_2}(z-z_3)^{2-a_3}}\ ,
\eea
subject to 
the following constraints imposed on the parameters
$$
a_1+a_2+a_3=2\ ,\ \ \ 
c_i=\frac{\delta_i+\delta_j-\delta_k}{z_j-z_i}+\frac{\delta_i+\delta_k-\delta_j}{z_k-z_i}\ ,\ \  \ \ \ (i,j,k)={\tt perm}(1,2,3)\ .
$$
In the case $\kappa=0$ the equation turns out to be  Riemann's differential equation.
By the simple change of variables and parameters, the 
ODE \eqref{sksaksa} can be reduced to the form  used in  Ref.\cite{Lukyanov:2012wq}.
Some particular cases of this equation were studied in a series of works
on integrable models of boundary interaction
\cite{Lukyanov:2003rt,Lukyanov:2003nj,Lukyanov:2006gv}.

Eq.\eqref{sksaksa}  was the starting point of the work \cite{BazU}. 
Bazhanov derived a system of integral  equations for  certain  connection coefficients
of the ODE \eqref{sksaksa}. It occurs to be identical
to Fateev's NLIE  taken at  the  CFT limit.
Needless to say that  the  limiting  form of the  NLIE
differs from the general one by
the   source terms only.
In all likelihood the original  Bazhanov derivation  can be adapted to the massive case.



\section*{Acknowledgments}

Numerous discussions with  V.V. Bazhanov, V.A. Fateev and A.B. Zamolodchikov were  highly valuable for me.
\bigskip

\noindent This  research was supported in part by DOE grant
$\#$DE-FG02-96 ER 40949.

\appendix
\section{Derivation of Eq.\eqref{ystasosposa} }

The purpose of this appendix is 
to rewrite the action functional \eqref{saoisosais} in terms of 
the field  $\eta$ and  coordinates  $(z,{\bar z})$.
We also outline the derivation of Eq.\eqref{ystasosposa}.

Under the conformal  map $w\to z$ the   image of  the arc $C_i$ enclosing  the vertex $w_i$
is  an
infinitesimal circle  of  radius  $\epsilon_i$,    related to the 
cut-off parameter $\epsilon$\ \eqref{sksosa} as
\bea\label{soisaisa}
\epsilon=
\frac{2\rho}{a_i}\ \bigg|\frac{z_{jk} }{z_{ij}  z_{ik} }\bigg|^{\frac{a_i}{2}}\ \epsilon^{\frac{a_i}{2}}_i\  .
\eea
Here $(i,j,k)$ stands for any  permutation of $(1,2,3)$ and  $|z_{ij}|=\sqrt{(z_i-z_j)({\bar z}_i-{\bar z}_j)}$.
One easily verifies the relation
\bea
\frac{1}{\pi\epsilon}\ \int_{C_i}{\rd \ell}\, {\hat \eta}=
a_i\, \eta_i
-\frac{a^2_i}{2}\ 
\log\bigg|\frac{z_{jk} }{z_{ij}  z_{ik} }\bigg|+
\frac{ a_i (2-a_i)}{2}\ \log(\epsilon_i)-a_i\ \log(\rho )
\eea
with
 $$\eta_i=\frac{1}{2\pi\epsilon_i}\ \oint_{|z-z_i|=\epsilon_i} \rd \ell \, \eta\ ,$$
and then
\bea
\label{jashja}
&&\frac{l_i}{\pi\epsilon}\ \int_{C_i}{\rd \ell}\, {\hat \eta}-
 a_i\,l_i^2\, \log(\epsilon)=2\, \big(\,m_i\,\eta_i
-m_i^2
\ \log(\epsilon_i)\,\big)+\frac{2-a_i}{2}\,\eta_i+\frac{(2-a_i)^2}{8}\       
 \log(\epsilon_i)\nonumber
\\
&&+ \Big( \frac{a_i^2-4}{8}-2m_i(m_i+1)\Big)
\ \log\bigg|\frac{z_{jk} }{z_{ij}  z_{ik} }\bigg|-
a_il_i (l_i+1)\ \log(\rho )+a_i\,l_i^2\, \log\Big(\frac{a_i}{2}\Big)\, .
\eea
The remaining necessary ingredient is 
\bea\label{ssisioa}
&& \lim_{\epsilon_i\to 0}
\,\frac{1}{\pi} \int_{|z-z_i|>\epsilon_i}
 {\rd^2 z}\
 \partial_z{\hat \eta}\partial_{\bar z}{\hat\eta}=-\frac{1}{8}\, \sum_{i=1}^3(a^2_i-4)\, \log\bigg|\frac{z_{jk} }{z_{ij}  z_{ik} }\bigg|+ \\
&&
\lim_{\epsilon_e\to 0\atop
R\to\infty}\bigg[\,
\,\frac{1}{\pi} \int_{|z-z_i|>\epsilon_i\atop
|z|<R}{\rd^2 z}\ 
 \partial_z{\eta}\partial_{\bar z}{\eta}-\sum_{i=1}^3\Big(\, \frac{2-a_i}{2}\ \eta_i
+\frac{(2-a_i)^2}{8}\ \log(\epsilon_i)\,\Big)+2\ \eta_{\infty}+2\,\log R\,\bigg]\ ,\nonumber
\eea
where
$$\eta_\infty=\frac{1}{2\pi R}\ \oint_{|z|=R}\rd \ell\,\eta\ .$$
Combining it with \eqref{jashja}, one arrives to Eq.\eqref{tystsaooiausais},
where
the constant ${\cal A}_0$ is given by
\bea\label{sospsaapsopa}
{\cal A}_0&=&
-\sum_{i=1}^32\,m_i(m_i+1)\, 
\log\bigg|\frac{z_{jk} }{z_{ij}  z_{ik} }\bigg|\\
&+&\sum_{i=1}^3\Big(\frac{1}{6}-\frac{(2m_i+1)^2}{a_i}\,\Big) \ \log(\rho)+
\sum_{i=1}^3a_i\,l_i^2\, \log\Big(\frac{a_i}{2}\Big)-
\frac{2}{\pi}\ A\ ,
\nonumber
\eea
and  $A$ is   the area of $\Sigma_{0,3}$
w.r.t. the flat metric \eqref{oiaoisao},
\bea\label{rssaosiosa}
A=\int\rd^2 z\ \sqrt{p(z){\bar p}({\bar z})}=\pi\rho^2\ 
\prod_{i=1}^3\gamma\Big({\frac{a_i}{2}}\Big)\ .
\eea

The  on-shell value, ${\cal A }_{\rm Liouv}^*$,  of the Liouville regularized action
\bea\label{sossapspaois}
{\cal A }_{\rm Liouv}[ \varphi]&=&\lim_{\epsilon\to 0\atop
R\to\infty}\, \bigg[\, \frac{1}{4\pi}\ \int_{|z-x_i|>\epsilon\atop
|z|<R}\rd^2 z\
\big(\, \partial_z{\varphi}\partial_{\bar z}{\varphi}+\re^{\varphi}\,\big)+
 \sum_{i=1}^3 \big(\,m_i\,\varphi_i-2\, m_i^2\,\log(\epsilon)\,\big)\nonumber\\
&+&2\,\log R+\varphi_\infty
\,\bigg]\ ,
\eea
where
$$\varphi_i=\frac{1}{2\pi\epsilon}\ \oint_{|z-z_i|=\epsilon}\rd \ell\,\varphi\, ,\ \ \ \ 
\varphi_\infty=\frac{1}{2\pi R}\ \oint_{|z|=R}\rd \ell\,\varphi\ ,$$
was calculated  in Ref.\cite{Zamolodchikov:1995aa}. The authors found
that the quantity
\bea\label{aosapsposa}
S^{{\rm (cl)}}=
{\cal A }_{\rm Liouv}^*+\sum_{i=1}^3\delta_i\, \log\bigg|\frac{z_{jk} }{z_{ij}  z_{ik} }\bigg|^2\ \ \ \ \ \ \ \ \ 
\big(\ \delta_i=-m_i(m_i+1)\ \big)
\eea
can be expressed in terms of the function $F$ \eqref{asosapsa} as\footnote{
\label{oias}See Eqs.(3.21),\,(4.12) in Ref.\cite{Zamolodchikov:1995aa}
where one should  set $\pi\mu b^2=1$ and $\eta_i$ is replaced by $(-m_i)$.}
\bea\label{asospas}
&&S^{{\rm (cl)}}=
F(-m_1-m_2-m_3-1)+F(-m_1-m_2+m_3)+
F(-m_2-m_3+m_1)\nonumber\\
&&+F(-m_3-m_1+m_2)-F(0)-F(-2m_1)-F(-2m_2)-F(-2m_3)\ .
\eea
Combining this result with \eqref{sospsaapsopa},
one founds
\bea\label{saooiausaisisus}
{\cal A}^*=
S^{{\rm (cl)}}+
\sum_{i=1}^3a_i\,l_i^2\, \log\Big(\frac{a_i}{2}\Big)
+\sum_{i=1}^3\Big(\frac{1}{6}-\frac{(2m_i+1)^2}{ a_i}\Big) \, \log(\rho)-\frac{2}{\pi}\ A
+o(\rho)
\eea
as $\rho \to 0$. Finally,   using the formula \eqref{apososa} for the constant ${\cal A}^*_\infty$,
one obtains Eq.\eqref{ystasosposa}.

\section{   Scalar potential for the stress-energy tensor}

Here we aim to obtain
some  useful identities  involving  ${\mathfrak q}_{1}$ and  ${\bar {\mathfrak q}}_{1}$.

The conserved charges   ${\mathfrak q}_{1}$ and  ${\bar {\mathfrak q}}_{1}$
can be expressed in terms of  the conventional  stress-energy tensor associated
with  the ShG equation
\bea\label{ospsaosa}
{\mathfrak q}_{1}= \re^{\frac{\ri\pi }{2}(a_1+a_2)}\ \oint_{C_w}\big(\,\rd w\ T+
\rd {\bar w}\ \Theta\,\big)\ ,\ \ \ \
{\bar {\mathfrak q}}_{1}=  \re^{\frac{\ri\pi }{2}(a_1+a_2)}\ \oint_{{\bar C}_w}\big(\,\rd {\bar w}\ {\bar T}+
\rd {\bar w}\ \Theta\,\big)\ .
\eea
Here  $T$, ${\bar T}$ and $\Theta$ stand for the non-vanishing
components  of the stress-energy tensor
\bea\label{sosisia}
T= (\partial_w{\hat \eta})^2\, ,\ \  \ \ \ \ \ T= (\partial_{\bar w}{\hat \eta})^2\ ,\ \ \ \ \ \Theta=4 \,\sinh^2({\hat \eta})\ .
\eea
The continuity equations
\bea\label{sa]assa}
\partial_{\bar w}T=\partial_w\Theta\ ,\ \  \ \ \ \partial_{ w} T=\partial_{\bar w}\Theta
\eea
imply that the fields \eqref{sosisia}
can be expressed in terms of a local potential function,
\bea\label{ystsossai}
T=\partial^2_w\Phi\ ,\ \ \  \ \  {\bar T}={ \partial}^2_{\bar w}\Phi\ ,\ \ \ \ \  \
\Theta=\partial_w\partial_{\bar w}\Phi\ ,
\eea
and therefore,
\bea\label{saopsospa}
{\mathfrak q}_{1}= \re^{\frac{\ri\pi }{2}(a_1+a_2)}\  
\oint_{C_w}(\rd w\,\partial_w+\rd {\bar w}\,\partial_{\bar w})\,\partial_w\Phi\ .
\eea
The integration contour here can be chosen to be
a union of  oriented segments as  shown in Fig.\ref{fig2a}.
It is straightforward to express the integral as a  sum of
the  discontinuities 
\bea\label{osapsap}
\Delta_1&=&2\ri\ \Big[\, \re^{\frac{\ri\pi a_1}{2}}\partial_w\Phi|_b-
\re^{-\frac{\ri\pi a_1}{2}}\partial_w\Phi|_{\bar b}\,\Big]\ \ \ \ \ {\rm as}\ \ \ b\in[w_1,w_3]\nonumber
\\
\Delta_2&=&2\ri\ \Big[\re^{-\frac{\ri\pi a_2}{2}}\partial_w\Phi|_a-\re^{\frac{\ri\pi a_2}{2}}\partial_w\Phi|_{\bar a}\,
\Big]\ \ \ \ \ {\rm as}\ \ \ 
a\in[w_2,w_3]\ .
\eea
Since
${\mathfrak q}_{1}$
does not actually depend on the location of the points
 $a,c\in[w_2,w_3]$ and  $b,d\in[w_1,w_3]$,  $\Delta_{1}$ and $\Delta_{2}$ 
remain constant along the sides $[w_2,w_3]$ and $[w_1,w_3]$, respectively. This yields the relation
\bea\label{ysopsaospsa}
{\mathfrak q}_{1}=
\sin\big({\textstyle \frac{\pi a_2}{2}}\big)\, \Delta_1+
\sin\big({\textstyle \frac{\pi a_1}{2}}\big)\, \Delta_2\, .
\eea
Similarly one finds 
\bea\label{ysopsaospsak}
{\bar {\mathfrak q}}_{1}=
\sin\big({\textstyle \frac{\pi a_2}{2}}\big)\, { \Delta}^*_1+
\sin\big({\textstyle \frac{\pi a_1}{2}}\big)\, { \Delta}^*_2\ .
\eea
Of course, in the  case under consideration $\Delta_{1}$ and $\Delta_{2}$ are real constants.

The formula \eqref{ysopsaospsa}
allows one to simplify the w.h.s. of Eq.\eqref{xsososa} from the main body of the paper.
Indeed,  the on-shell value of  the  trace of the stress-energy tensor
is  given by the Laplacian of the  scalar potential $\Phi$\ \eqref{ystsossai}. Hence,
\bea\label{isxssosa}
\rho\  \frac{\partial  {\mathfrak Y}}{\partial \rho}=
 \lim_{\epsilon\to 0}\,  \frac{2}{\pi}\int_{D^{\rm (reg)}} \rd^2 w \  \partial_w\partial_{\bar w}\Phi
-\sum_{i=1}^3
a_i\, l^2_i\ .
\eea
As it follows from  the boundary conditions \eqref{osasail},
\bea\label{usyaosasaposaso}
\Phi(w,{\bar w})=-2l_i^2\ \log|w-w_i|+O(1)\ \ \ \ \ {\rm at}\ \ \ \ w\to w_i\ .
\eea
Combine this relation with  the above  observation  that the discontinuities $\Delta_{1}$ and $\Delta_2$ \eqref{osapsap}
remain constant along the corresponding sides of the polygon $(w_1,w_3,w_2,{\bar w}_3)$, one finds 
\bea\label{aposussaytsrs}
\rho\  \frac{\partial  {\mathfrak Y}}{\partial \rho}=-\frac{1}{2\pi}\ (\,\Delta_1\,|w_3-w_1|+\Delta_2\,|w_3-w_1|\,)\ .
\eea
Using  \eqref{ysopsaospsa}, 
the w.h.s. of \eqref{aposussaytsrs} can be rewritten in terms of the
conserved charge  ${\mathfrak q}_1={\bar {\mathfrak q}}_1$ and
 the circumdiameter $d$ of the triangle   $ (w_1,w_2,w_3)$ (see  Fig.\,\ref{fig8} and Eq.\eqref{soosapsosa}).
This yields   Eq.\eqref{sissisusaopsopsaps}.

\section{Reflection amplitude}

Here we discuss the variations of the Yang-Yang function with respect to the parameters $l_i$.

Varying     the   action \eqref{saoisosais} with the use of the on-shell
condition
$\delta {\cal A}|_{l_i,\rho,a_i-{\rm fixed}}=0$,
one easily derive the relation
\bea\label{aosapass}
\delta_l {\cal A}^*=a_i\, {\hat \eta}_i\ \delta l_i\ ,
\eea
where the constants ${\hat \eta}_i$
can be regarded as regularized 
values of the
solution ${\hat\eta}$ at the boundary $\partial {\mathbb D}$:
\bea\label{aaosisoasi}
 {\hat \eta}_i=\lim_{|w-w_i|\to 0}\big({\hat\eta}(w,{\bar w})-2\,l_i\ \log|w-w_i|\,\big)\ .
\eea
It should be stressed that unlike  $l_i$,  which are   ``input'' parameters applied
with the problem, the values of the constants  ${\hat \eta}_i$ are  not prescribed in advance
but determined through the solution, i.e. it is rather part
of the ``output''.

Let us  define
\bea\label{sossaoa}
{\mathfrak S}_i=2^{\frac{8p_i}{a_i}-2}\ \gamma\Big({ \frac{2 p_i}{a_i}}\Big)\ \re^{2 {\hat \eta}_i}\ ,
\eea
where we use the notation
\bea\label{asospospasasp}
p_i=m_i+\frac{1}{2}=\frac{a_i}{2}\ \Big(l_i+\frac{1}{2}\Big)\ .
\eea
Bearing in mind the definition of the Yang-Yang function, \eqref{aosapass} can be rewritten in the form
\bea\label{sosopsaposa}
\frac{\partial {\mathfrak Y} }{\partial p_i}=\log {\mathfrak S}_i\ .
\eea
As  it follows from the small-$\rho$ expansion \eqref{asosposa},
\bea\label{sospssa}
{\mathfrak S}_i= S(p_i|p_j+ p_k) S(p_i|p_j-p_k)\ \exp\Big(\sum_{n=1}^\infty \frac{\partial y_n}{\partial p_i}\  \rho^{4n}\,
\Big)\ .
\eea
Here $(i,j,k)$ stands for  any  permutation of $(1,2,3)$ and
\bea\label{spssaios}
S(p_i|q)=\bigg(\frac{\rho}{a_i}\bigg)^{-\frac{2p_i}{a_i}}\
\frac{\Gamma(\frac{1}{2}+p_i+q) \Gamma(\frac{1}{2}+p_i-q)}{\Gamma(\frac{1}{2}-p_i+q)
\Gamma(\frac{1}{2}-p_i-q)}\
\frac{\Gamma(1-2 p_i)}{\Gamma(1+2 p_i)}\
\frac{\Gamma(1+ \frac{2 p_i}{a_i})}{\Gamma(1-\frac{2 p_i}{a_i}) }\ .
\eea
Notice that,   with  the parameter identifications \eqref{sssaopsa},\,\eqref{saossaops},
\bea\label{asopospaos}
\Big(\frac{2\pi}{R}\Big)^{{\frac{2p_i}{a_i}}}\, S(p_i|q)
\eea
coincides with the so-called ``reflection amplitude''-- an important characteristic of the Fateev model
(see Ref.\cite{Baseilhac:1998eq} for details).

\end{document}